\documentclass[11pt,a4paper]{article}
\pdfoutput=1
\usepackage{jcappub}
\usepackage{bm}
\usepackage{empheq}

\newcommand{\fnl}{f_{\mathrm{NL}}}

\newcommand{\be}{\begin{equation}}
\newcommand{\ee}{\end{equation}}
\newcommand{\bea}{\begin{eqnarray}}
\newcommand{\eea}{\end{eqnarray}}
\newcommand{\eref}{\eqref}

\newcommand{\vect}[1]{\bm{\mathrm{{#1}}}}

\graphicspath{{./}}
\notoc

\usepackage[T1]{fontenc} 

\usepackage{braket}

\numberwithin{equation}{subsection}

\newcommand\ben{\begin{enumerate}}
\newcommand\een{\end{enumerate}}
\newcommand\bal{\begin{align*}}
\newcommand\eal{\end{align*}}
\newcommand\bi{\begin{itemize}}
\newcommand\ei{\end{itemize}}

\def\t{\text}

\def\id{\protect{{1 \kern-.28em {\rm l}}}}

\newcommand{\gae}{\lower 2pt \hbox{$\, \buildrel {\scriptstyle >}\over {\scriptstyle
\sim}\,$}}
\newcommand{\lae}{\lower 2pt \hbox{$\, \buildrel {\scriptstyle <}\over {\scriptstyle
\sim}\,$}}

\usepackage{graphicx}
\usepackage{caption}
\usepackage{subcaption}

\begin{document}

\title{The squeezed limit of the bispectrum in multi-field inflation}

\author{Zachary Kenton}
\emailAdd{z.a.kenton@qmul.ac.uk}
\author{and David J. Mulryne}
\emailAdd{d.mulryne@qmul.ac.uk}

\affiliation{School of Physics and Astronomy, Queen Mary University of London,\\
Mile End Road, London, E1 4NS,  UK.}

\date{\today}

\abstract{
We calculate the squeezed limit of the bispectrum produced by
inflation with multiple light fields. To achieve this we allow for different
horizon exit times for each mode and calculate the intrinsic field-space three-point function 
in the squeezed limit using soft-limit techniques. We then use the $\delta N$ formalism from the time the last mode exits the horizon 
to calculate the bispectrum of the primordial curvature perturbation.
 We apply our results to calculate the spectral index of the halo bias, $n_{\delta b}$, 
an important observational probe of the squeezed limit of the primordial bispectrum and compare 
our results with previous formulae.
We give an example of a curvaton model with   
$n_{\delta b} \sim {\cal O}(n_s-1)$ 
for which we find a 
20\% correction to observable parameters
for squeezings relevant to future experiments.
For completeness, we also calculate the squeezed 
limit of three-point correlation functions involving gravitons for multiple field models. }


\maketitle
\newpage
\tableofcontents
\section{Introduction}

Reliable calculations of the N-point correlation functions of the primordial
curvature perturbation, $\zeta$, are essential to confront models of inflation with 
present and future observational constraints.  
Since inflation can occur at energies as high as $10^{14}\text{GeV}$, these 
correlation functions provide an unparalleled observational window into high energy physics, 
providing information about the fields and their interactions active in the early universe. 
In this paper we consider the three-point correlation function, and discuss how 
it can be calculated in a particular limit, known as the `squeezed limit'. 
Our primary  aim 
is to provide clarity on how to accurately confront models of inflation with more 
than one light field 
against observations sensitive to this limit. This includes models in 
which more than one field supports inflation, as well as spectator models such 
as the curvaton scenario and modulated reheating models -- we refer to such models as multi-field models, and exclude 
cases in which additional heavy fields play a role.

The three-point function, parametrised by the bispectrum, is a function of three 
wave vectors which sum to zero as a result of momentum conservation, forming 
 a triangle in momentum space. The squeezed limit  
refers to the case where one of the associated 
wave numbers is much smaller than the other two, such that the 
triangle looks `squeezed'.
Calculations of the three-point function are now extremely mature, yet 
for technical reasons previous multiple field calculations have only been been performed 
explicitly for the case of a mild hierarchy between wave numbers, as we will see. The 
highly squeezed limit is, however, very important both from an observational and a 
theoretical point of view. 

The squeezed limit of the the bispectrum is the simplest possible example of a more general class of 
limits of correlation functions, referred to as soft limits. Soft limits occur when 
there exists a separation of scales in a physical problem. 
In the inflationary context, 
soft limits of correlation functions of the primordial curvature perturbation 
offer an exciting opportunity to confront theory with  
observations. For example, in the case of a single slow-roll field with canonical kinetic terms, Maldacena \cite{Maldacena:2002vr} found that the bispectrum of the curvature perturbation in the squeezed limit is purely determined by the tilt of the power spectrum, with the assumption of a Bunch-Davies initial state. The relation is: $12\fnl^{\rm sq} =-5(n_s-1)$, 
where $\fnl^{\rm sq} $ is the squeezed limit of the 
reduced bispectrum, and $n_s-1=-0.032 \pm 0.006$ \cite{Ade:2015lrj} is the spectral tilt. Creminelli \& Zaldarriaga \cite{Creminelli:2004yq} (see also Ref.~\cite{2011JCAP...11..038C}), showed that Maldacena's result holds even without the assumption of slow-roll\footnote{See, for example, \cite{Khoury:2008wj,2012JCAP...05..037R} for single field, but non-slow-roll models which obey Maldacena's relation.} in all models with a Bunch-Davies initial state and where the classical solution is a dynamical attractor  -- the proof of which was later formalized by Cheung {\it et al.} \cite{Cheung:2007sv}. Thus a detection of $\fnl^{\rm sq} \gae {\cal O}(0.01)$ would rule out all single field models with a Bunch-Davies initial state and where the classical solution is a dynamical attractor\footnote{See \cite{Berezhiani:2014kga} (and e.g. \cite{Ashoorioon:2013eia}) where more general initial states are considered, and see \cite{Mooij:2015yka} (and references therein) where non-attractor models are considered.}. 
For single field inflation, considerable work has also gone into studying 
more general soft limit results and providing further consistency relations amongst correlation functions \cite{PhysRevD.74.121301,Mooij:2015yka,Li:2008gg,Leblond:2010yq,Seery:2008ax,Hinterbichler:2012nm,
Creminelli:2012ed,McFadden:2014nta,Joyce:2014aqa,Mirbabayi:2014zpa,Hinterbichler:2013dpa,
Weinberg:2003sw,Senatore:2012wy,Goldberger:2013rsa,Weinberg:2005vy,Senatore:2009cf,
Flauger:2013hra,Pimentel:2013gza,Berezhiani:2014tda,Tanaka:2011aj,Pajer:2013ana,
Creminelli:2011sq,Berezhiani:2014kga}. 
Moreover, soft limits can be used to provide information about other fields present during inflation \cite{Assassi:2012zq,Kehagias:2015jha,Arkani-Hamed:2015bza,Mirbabayi:2015hva,Sugiyama:2011jt,Sugiyama:2012tr,PhysRevLett.107.191301,PhysRevD.77.023505}.

In the case of inflation with multiple light fields, in contrast to the single field case, {\it model independent} results such as the Maldacena consistency relation are not 
possible. However, many observables which constrain  
non-Gaussianity produced by multiple field inflation are particularly sensitive to soft limits. Examples include\footnote{One might naively think that the long-wavelength fluctuations used to model the observed CMB power asymmetry (see e.g. \cite{Erickcek:2008sm,Erickcek:2008jp,Kobayashi:2015qma,Byrnes:2015asa,Kenton:2015jga}) may also be an observational probe of soft limits  -- however, to describe the asymmetry, the long wavelength mode is required to be superhorizon, and so won't be suitable as a soft momentum in the correlation functions considered in this paper.} the spectral index of the halo bias \cite{PhysRevD.77.123514} and CMB $\mu$-distortions \cite{2012PhRvL.109b1302P}. This is why explicit calculations of soft limits in the multiple field context are important when comparing {\it model  dependent} predictions against observation.

In this paper, as a starting point to more general studies of soft limits in multiple field inflation, we explore the simplest case of the squeezed limit of  the bispectrum. In most previous studies, the path to the bispectrum for multiple field models has been to first 
use the in-in formalism to calculate the three-point function of scalar field perturbations at a time soon after all modes have left the horizon, as was first done by Seery \& Lidsey \cite{Seery:2005gb}. Next the $\delta N$ formalism \cite{Sasaki1995,Wands2000} is applied to convert the field-space correlations to correlations of $\zeta$ \cite{Lyth2005}. The result of Seery \& Lidsey for the three-point function, however, requires that there is not a large hierarchy between the three wavenumbers involved in the bispectrum, and thus that  the modes of the bispectrum must cross the horizon during inflation at roughly the same time. As is clearly stated in their paper, therefore, their result is not valid in the highly squeezed limit where there is an appreciable difference in the exit times of different modes. Moreover, on using $\delta N$ to convert from field-space fluctuations to $\zeta$, one finds that the three-point function of the curvature perturbation involves copies of the two-point function of field fluctuations evaluated after all modes have exited the horizon. 
At this point in the procedure, previous explicit calculations have considered at most a mild hierarchy between the scales at which these two-point functions are evaluated \cite{2010JCAP...02..034B,Dias:2013rla}, with $|\log(k_1/k_3)| \sim \mathcal{O}(1)$, where $k_1$ is the long-wavelength mode, and $k_3$ is the short-wavelength mode.

Future experiments, together with the expected amount of squeezing they will be sensitive to, are shown in Table~\ref{tab:exp} \cite{Dias:2013rla}. These experiments will probe a hierarchy much larger than that allowed by  previous theoretical calculations. Therefore, it is very important to have theoretical predictions for multiple field models valid in the highly squeezed limit, to be able to compare  with observations. 

\begin{table}[!h]
\centering
\begin{tabular}{|c|c|c|c|}
\hline 
Experiment & Dark Energy Survey & Euclid & $\mu$-distortions \\ 
\hline 
Squeezing & $\log(k_1/k_3) \sim -2$ & $\log(k_1/k_3) \sim -8$ & $\log(k_1/k_3) \sim -19$ \\ 
\hline 
\end{tabular} 
\caption{Future experiments and their observable range of scales.}
\label{tab:exp}
\end{table}

In our study, therefore, we wish to relax the requirement of a mild hierarchy, 
and study models explicitly in the highly squeezed limit. To do so we will 
use a similar soft limit argument to Cheung {\it et al.} \cite{Cheung:2007sv}, but applied 
to calculate the 
three-point function of the scalar field perturbations. This result 
reduces to that of Seery \& Lidsey for mild squeezing, but is valid in 
the highly squeezed limit and does not rely on slow-roll. In analogy with the single field case, where the three-point function depends on the tilt of the power spectrum,  we find that the three-point function of the scalar field perturbations 
 depends on derivatives of the field space two-point  function with respect to the 
background value of the fields. 
Armed with this result, we further relax the usual assumption 
that the copies of the two-point function of the field perturbations, 
which appear in the three-point function for $\zeta$, involve only a mild hierarchy 
of scales. This can easily be achieved by accounting for the evolution between horizon 
crossing times. We find that for any model in which previous applications of the $\delta N$ 
formalism give rise to analytic results, analytic expressions for the highly squeezed limit are also possible. 
We finish by considering a specific curvaton model and compare our new squeezed limit 
formulae with previous expressions. We find significant differences for cases in which non-Gaussianity 
depends on scale.

The outline of this paper is as follows: in \S\ref{reviewdn} we review the previous applications of the $\delta N$ formalism, taking a pedagogical approach in this section which will be helpful when we come 
to extend previous work. In \S\ref{sec:deltaNdiffcross} we conveniently parametrise the evolution of the superhorizon field perturbations between exit times in terms of a `$\Gamma$-matrix', and calculate the three-point function of the scalar field perturbations at the time the last mode exits the horizon using soft-limit arguments. 
Putting these elements into the $\delta N$ formalism we calculate the highly squeezed limit of the bispectrum of the curvature perturbation. In \S\ref{sec:scaledep} we calculate the scale dependence of the squeezed limit of the reduced bispectrum, focussing on the spectral index of the halo bias. In \S\ref{sec:gammaimport} we provide explicit formulae for the $\Gamma$ matrix, and 
investigate the concrete example of the mixed inflaton-curvaton scenario \cite{2012JCAP...06..028F}  with self-interactions \cite{Byrnes:2015asa}.

Throughout this paper we work in units where $\hbar = c= 1$ and we set the reduced Planck mass $M_p=1$.

\section{Review of $\delta N$ } \label{reviewdn}
In this section we review previous applications of the 
$\delta N$ formalism to calculate the bispectrum of  $\zeta$. In doing so, we present 
key definitions and formulae essential for the subsequent sections.

\subsection{Defining $\zeta$}
We will be considering Fourier modes 
which make up a given triangle of the bispectrum, labelled 
by three wave vectors 
$\vect{k_1}$, $\vect{k_2}$ and $\vect{k_3}$. These 
scales satisfy the condition $\vect{k_1}+\vect{k_2}+\vect{k_3}=0$ because 
of momentum conservation, and without loss of generality we order them 
such that $k_1\leq k_2\leq k_3$, 
where $k = |\vect{k}|$. 
We denote the times these modes cross the horizon by $\{t_1, t_2,t_3\}$,   
defined by the condition $k_1 = a(t_1)H(t_1)$ etc. and so we have $t_1 \leq t_2 \leq t_3$. For the bispectrum there are only two 
limiting possibilies for the relative magnitudes of wave numbers. The close to the equilateral case, 
with $k_1\approx k_2 \approx k_3$, and the  squeezed shape, with $k_1\ll k_2\approx k_3$. The amount of squeezing 
can be measured by the value of $\log (k_1/k_3)$, a negative number with greater absolute value 
indicating greater squeezing.

In the context of multiple field inflation, the cosmological perturbation of most interest 
is the uniform-density curvature perturbation, which coincides with the 
comoving curvature perturbation on superhorizon scales, where it is conserved if the system reaches an adiabatic limit \cite{Lyth2004,Rigopoulos:2003ak}. 
In general the curvature perturbation can be defined as a
scalar perturbation to the spatial metric  for a given foliation of spacetime, 
and moreover
can be written as a local perturbation to 
the scale factor as \cite{Lyth:2005du,Lyth2004,PhysRevD.42.3936}
\begin{align}
a(t,\vect{x})= a(t) e^{\psi(t,\vect{x})}.
\end{align}
Choosing the $t$-slicing to be such that the spatial hyper surfaces have 
uniform-density (UD) leads to $\psi_{\text{UD}}(t,\vect{x})\equiv\zeta(t,\vect{x})$, while 
a flat $t$-slice is defined by $\psi_{\text{flat}}(t,\vect{x})=0$. The power 
spectrum, $P_{\zeta}$, and bispectrum, $B_{\zeta}$, of the curvature 
perturbation $\zeta$ are then defined by the two and three-point correlation functions
\begin{align}
\langle \zeta_{\vect{k_1}}\zeta_{\vect{k_2}} \rangle &= P_{\zeta}(k_1)  (2 \pi)^3 \delta(\vect{k_1} + \vect{k_2})
\\
\langle \zeta_{\vect{k_1}}\zeta_{\vect{k_2}}\zeta_{\vect{k_3}} \rangle &=  B_{\zeta}(k_1,k_2,k_3)(2 \pi)^3  \delta(\vect{k_1} + \vect{k_2} + \vect{k_3}).
\end{align}

\subsection{The $\delta N$ formalism}
Using the spatially dependent definition of the scale factor, the  
number of e-folds which occurs between two time 
slices of the perturbed spacetime, labeled by $T$ and $t_u$ respectively, is 
a function of position and is given by 
\begin{align}
 N(T,t_u,\vect{x}) \equiv \int_{T}^{t_u}\frac{\dot{a}(t,\vect{x})}{a(t,\vect{x})}dt = \int_{T}^{t_u}H(t)dt + \psi(t_u,\vect{x}) - \psi(T,\vect{x})
\end{align}
while the unperturbed number of e-folds is given by $ N_0 (T,t_u) \equiv \int_{T}^{t_u}H(t)dt$. Taking the $T$-slices to be flat, and the $t_u$-slices to be uniform density gives
\begin{align}
\zeta(t_u,\vect{x}) =  N(T, t_u,\vect{x}) -  N_0 (T, t_u) \equiv \delta N(t_u,\vect{x})\label{deltaNdef}
\end{align}
which is the celebrated $\delta N$ formula \cite{Sasaki1995}. We note that 
$\delta N$ doesn't  depend on the initial time $T$ \cite{Lyth2004}. Typically 
$T$ is taken to be some time after all the modes involved in a given correlation function 
have exited the horizon (for the bispectrum this means $T>t_3$). If the system becomes adiabatic $\zeta$ is also independent of $t_u$ and so the later time $t_u$ should be the time at which adiabaticity is reached. If it doesn't become adiabatic, then $t_u$ can be taken to be the time at which the correlations are required (see for example the discussion in Ref.~\cite{2011JCAP...11..005E}). 

To use this formalism in practice, we must employ 
the separate universe approximation \cite{Lyth:1984gv,Wands2000} to cosmological perturbation theory. 
This states that on super-horizon scales positions in the 
perturbed universe evolve independently of one another, and 
do so according to the same equations as the unperturbed cosmology, so that every 
position can be treated as a `separate universe'. The number of e-folds which occur at every position can therefore be calculated using the local conservation and  Friedmann equations in that separate universe.

For an inflationary model with $n$ scalar fields, $\phi_i$, where $i$ runs from 
one to $n$,  
we can split the field values on any flat slice into background and perturbed parts 
$\phi_i(t,\vect{x})= \phi_i(t)+ \delta \phi_i(t,\vect{x})$. 
Further demanding that the slow-roll equations of motion 
are satisfied at time $T$, such that 
$3H^2 =V(\phi_i)$ and $3H\dot{\phi}_{i} = -V_{,i}$ , the initial 
conditions at time $T$ for the 
perturbed cosmology become 
dependent only on $\phi_i(T, \vect{x})$. Consequently, the number 
of e-folds to any subsequent time slice becomes a function only of the initial field values, even if the cosmology evolves away from slow-roll.  
In particular, we can write
\begin{align}
  N(T,t_u,\vect{x})= N((\phi_i(T,\vect{x})), t_u)=N((\phi_i(T)+ \delta \phi_i(T,\vect{x})), t_u)
\end{align}
which gives
\begin{align}
\zeta(\vec{x}) =  N((\phi_i(T)+ \delta \phi_i(T,\vect{x})), t_u) -  N_0(\phi_i(T), t_u).
\end{align}
One can then Taylor expand in the initial flat slicing field perturbations, which in Fourier space 
leads to \cite{Sasaki1995,Lyth2005}
\begin{align}
\zeta_{\vect{k_1}} &=  N^{(T)}_{i } \delta \phi^{(T)}_{i,\vect{k_1}} + \frac{1}{2}N^{(T)}_{ij}( \delta \phi^{(T)}_{i } \star\delta \phi^{(T)}_{j})_{\vect{k_1}}+...
\label{zetaatstar}
\\
\text{where } N^{(T)}_{i} &\equiv \frac{\partial N_0}{\partial \phi^{(T)}_{i }} \label{nicoeff}\,,
\end{align}
and where $\star$ denotes convolution. The additional vector (boldface) subscript indicates wavevector and the superscript in brackets is a shorthand indicating evaluation time
\begin{align}
\phi_i^{(T)}\equiv \phi_i(T)
\end{align}
which we use from now on. 

The correlation functions of the field perturbations are defined as
\begin{align}
\langle \delta \phi^{(T)}_{i,\vect{k_1}}  \delta \phi^{(T)}_{j,\vect{k_2}}  \rangle 
&=
 \Sigma_{ij}^{(T)}({k_1})   (2 \pi)^3 \delta(\vect{k_1} + \vect{k_2}) \label{twopointdeltaphi}
\\
\begin{split}
\langle  \delta \phi^{(T)}_{i,\vect{k_1}}\delta \phi^{(T)}_{j,\vect{k_2}}\delta \phi^{(T)}_{k,\vect{k_3}}\rangle 
&=  \alpha_{ijk}^{(T)}(k_1,k_2,k_3)
 (2 \pi)^3 
  \delta(\vect{k_1} + \vect{k_2} + \vect{k_3})
\end{split} 
\label{eq:fieldBi}
\end{align}
such that the power spectrum \cite{Sasaki1995} and 
bispectrum \cite{Lyth2005} of $\zeta$ are given by 
\begin{align}
P_{\zeta}(k_a) = &N^{(T)}_{i }N^{(T)}_{j }\Sigma_{ij}^{(T)}({k_a})   \label{powerdeltaN}
\\
\begin{split}
B_{\zeta}(k_1,k_2,k_3) = &N^{(T)}_{i }N^{(T)}_{j }N^{(T)}_{k }\alpha_{ijk}^{(T)}(k_1,k_2,k_3)
\\
&+N^{(T)}_{i }N^{(T)}_{jk }N^{(T)}_{l }  \big [\Sigma_{ij}^{(T)}({k_1})\Sigma_{kl}^{(T)}({k_2}) 
+ (k_1 \to k_2 \to k_3  )\big ]
\end{split} \label{bispectrumdeltaN}
\end{align}
where the arrows indicate there are two additional terms formed by cyclic permutations. 

One can then define the reduced bispectrum, conventionally denoted by $\fnl(k_1,k_2,k_3)$, by 
comparing the amplitude of the bispectrum to the square of the power spectrum as follows
\begin{align}
\fnl(k_1,k_2,k_3) \equiv \frac{5}{6}\frac{B_{\zeta}(k_1,k_2,k_3)}{[P_{\zeta}(k_1)P_{\zeta}(k_2)+ (k_1 \to k_2 \to k_3  )]}.\label{fnl}
\end{align}

\subsection{$\delta N$ for $k_1 \approx k_2 \approx k_3$}\label{subsec:deltaN}

For light, canonically normalized fields, the field-space 
correlation function for a given wavenumber takes a very simple 
form at the time the wavenumber crosses the horizon\footnote{
Strictly speaking this result is the form the two-point function takes 
once the decaying mode present at horizon 
crossing has been lost, written in terms of horizon crossing 
parameters.} 
(and from which the $\delta N$ formalism can be employed), one finds \cite{Stewart:1993bc,Nakamura:1996da}
\begin{align}
\label{canonicalsigma}
\Sigma_{ij}^{(1)}(k_1) = \frac{{H^{(1)}}^2}{2 k_1^3}\delta_{ij}.
\end{align}
In the most common application of $\delta N$, one assumes a 
near-equilateral momentum regime, where all the wavenumbers are approximately 
equal, $k_1 \approx k_2 \approx k_3$ and thus 
the horizon crossing times of the three wavenumbers involved in the bispectrum 
can be identified with a single time, $t_*$, such that 
$t_1 \approx t_2 \approx t_3 \approx t_*$.
In this regime, it is then common to make the simple choice 
$T=t_*$.

For canonical slow-roll inflation, one can 
take\footnote{The leading order in slow-roll correction to this \cite{Stewart:1993bc,Nakamura:1996da,2013JCAP...10..062D} is to replace $\delta_{ij} \mapsto \delta_{ij} +2cu_{ij}$, where $c\equiv 2-\log 2 - \gamma$, with $\gamma$ the Euler-Masheroni constant, 
and $u_{ij}=-(\log V)_{,ij}$, but we will not use this correction in this work. \label{foot:sigma}} 
\begin{align}
\label{canonicalsigma2}
\Sigma_{ij}^{(*)}(k_1) \approx \frac{{H^{(*)}}^2}{2 k_1^3}\delta_{ij}\,,
\end{align}
and similarly for the other wavenumbers, $k_2$ and $k_3$. 
Moreover, Seery \& Lidsey \cite{Seery:2005gb} used the in-in formalism to calculate $\alpha_{ijk}^{(*)}(k_1,k_2,k_3)$ in the close to equilateral momentum configuration at the time $t_*$. They found
\begin{align}
\begin{split}
\alpha_{ijk}^{(*)}(k_1,k_2,k_3) &= \frac{4\pi^4}{k_1^3k_2^3k_3^3}\left (\frac{{H^{(*)}}}{2\pi}\right )^4\sum_{\text{6 perms}}\frac{\dot{\phi}_i^{(*)}\delta_{jk}}{4{H^{(*)}}}\left (-3\frac{k_2^2k_3^2}{k_t} - \frac{k_2^2k_3^2}{k_t^2}\left (k_1+2k_3\right ) +\frac{1}{2}k_1^2 - k_1k_2^2 \right )
\end{split} \label{SLalpha}
\end{align}
where the sum is over the six permutations of $(ijk)$ while simultaneously rearranging the momenta $k_1,k_2,k_3$ such that the relative positioning of the $k$'s is respected. We emphasize that this result assumed that the three $k$ modes crossed the horizon at roughly the same time and cannot be trusted when the crossing times are too different\footnote{This is because in evaluating the time integrals of the in-in calculation, the time-dependent coefficients of the field perturbations were all evaluated at the common horizon crossing time, $t_*$, of the near-equilateral modes. This can't be done if the modes exit at largely different times, as is the case in the highly squeezed limit.}, as in the case of the highly squeezed limit, which we will consider in this paper.

Sticking with the near-equilateral configuration, with $k_1 \approx k_2 \approx k_3$, and taking the $\delta N$ initial time $T=t_*$, the results \eqref{canonicalsigma} and \eqref{SLalpha} can then be used together with Eqs.~(\ref{powerdeltaN})-(\ref{fnl}) to give the well-known expression \cite{Lyth2005}, valid when 
$\fnl(k_*) \gg {\cal O}(\epsilon)$,
\begin{align}
\fnl \approx \frac{5}{6}\frac{N^{(*)}_{i}N^{(*)}_{ij}N^{(*)}_{j}}{(N^{(*)}_{l}N^{(*)}_{l})^2} \,,
\label{eq:fnl}
\end{align}
where \eqref{SLalpha} has been used to demonstrate that  
the second term in Eq.~(\ref{bispectrumdeltaN}) must dominate over the first term if $\fnl$ is to be large -- i.e. the contribution from $\alpha$ is neglected \cite{Lyth:2005qj}.
Eq.~(\ref{eq:fnl}) retains a dependence on $k_*$ through the 
horizon crossing time $t_*$, and the relation that $k_*=a(t_*)H(t_*)$. 

Given that it retains this dependence on the crossing scale, which is a 
measure of the overall scale of the bispectrum in the near-equilateral configuration, the bispectrum 
calculated above is sometimes referred to as quasi-local (to be contrasted with the local shape, in which $\fnl$ is independent of all three $k$'s). Differentiating \eqref{eq:fnl}, Byrnes {\it et al.} \cite{2010JCAP...02..034B} gave an expression for the tilt of $\fnl$, denoted $n_{\fnl}$, 
for equilateral triangles
\begin{align}
n^{(*)}_{\fnl } &\equiv \frac{\mathrm{d} \log |\fnl|}{\mathrm{d} \log k} = \frac{(n^{(*)}_{i}+ n^{(*)}_{j} + n^{(*)}_{ij})N^{(*)}_{i}N^{(*)}_{ij}N^{(*)}_{j}}{N^{(*)}_{l}N^{(*)}_{lm}N^{(*)}_{m}} - 4\hat{n} \label{nfnleq}
\\
\text{where } n^{(*)}_{i} &\equiv \frac{1}{H}\frac{\mathrm{d} \log N^{(*)}_{i}}{\mathrm{d} t_*}, \qquad n^{(*)}_{ij} \equiv \frac{1}{H}\frac{\mathrm{d} \log N^{(*)}_{ij}}{\mathrm{d} t_*}
\qquad
\text{ and }  \hat{n} \equiv  \frac{   n^{(*)}_{i}N^{(*)}_{i}N^{(*)}_{i}}{N^{(*)}_{j}N^{(*)}_{j}}.
\end{align}

\subsection{Beyond the near-equilateral configurations}

In the present paper we aim to be go beyond the case where 
$k_1 \approx k_2 \approx k_3$, to study the highly squeezed limit. Before 
we do, we must stress that although most previous studies use the formulae presented 
in the previous sub-section, some attempts have already been made to relax this 
near-equilateral assumption, which 
we briefly review now. Full details will appear when we compare our expressions to earlier ones in 
later sections. 

In Ref.~\cite{2010JCAP...02..034B},  in addition to the consideration 
of scale dependence discussed above, Byrnes {\it et al.} considered the 
more general dependence of the reduced bispectrum on $k_1$, $k_2$ and $k_3$, using the 
$\delta N$ framework. They did so 
by solving the slow-roll field equations for scalar fluctuations to first order in the quantity
$|\log(k_1/k_3)|$  about the time $t_*$. This was used to allow a 
mild hierarchy of scales. The result becomes inaccurate as this 
quantity grows with increased squeezing. Moreover, they did not calculate the squeezed limit of $\alpha_{ijk}$, assuming it to contribute negligibly for an observable bispectrum of $\zeta$ -- as is the case for near-equilateral configurations.
We give more details on their approach and when it is accurate 
in \S\ref{sec:tilt} and Appendix~\ref{App:BNTW}. 

In Ref.~\cite{Dias:2013rla} Dias {\it et al.} took a different approach. They used a next to leading 
order expression in slow-roll for the bispectrum at the time $t_*$, 
calculated by the same authors in Ref.~\cite{2013JCAP...10..062D}.  Such next 
to leading order expressions  automatically provides 
information on scale dependence, and allows for a mild hierarchy of scales. 
They went on to use their expressions for an investigation of the spectral index of the halo bias.
The Dias {\it et al.} result includes the form of $\alpha_{ijk}$ in the case of a 
mild hierarchy, but also takes the form of an expansion in $|\log(k_1/k_3)|$ 
with only the first term calculated, 
and so again becomes inaccurate as the squeezing is increased. Their result is given 
explicitly in Appendix~\ref{App:DRSreduction}. 

A major aim of this paper is to determine when previous expressions can be trusted, 
and when the squeezing becomes large enough that the expressions we present are 
necessary. 

\section{The squeezed limit of the bispectrum with $\delta N$} \label{sec:deltaNdiffcross}

In this paper, therefore, we wish to extend previous work to explicitly calculate the bispectrum at 
time $t_u$ in the case of a truly squeezed momentum 
configuration, $k_1 \ll k_2 \approx k_3$, which will involve (perhaps very) 
different horizon exit times $t_{1}\ll t_{2}\approx t_{3}$. 
To do so we will 
employ expression \eqref{bispectrumdeltaN}, setting $T=t_3$, 
but will fully account for the different horizon crossing times, and 
also show how to calculate  $\alpha_{ijk}^{(3)}$ for highly squeezed configurations.
In contrast to the previous section, 
the new objects we need to calculate are $\Sigma_{ij}^{(3)}({k_1})$, 
the field perturbation two-point function for modes which cross the horizon at time $t_1$ 
evaluated at the later time $t_3$, and $\alpha_{ijk}^{(3)}(k_1,k_2,k_3)$ for the $k_1 \ll k_2 \approx k_3$ 
configuration. Many authors 
have discussed how to propagate the field-space two-point correlation function past horizon crossing \cite{Stewart:1993bc,Nakamura:1996da,2013JCAP...10..062D}, but to the best of our knowledge no one has then 
employed these techniques to explicitly investigate the bispectrum of $\zeta$ 
beyond cases of mild squeezing. Moreover, to the best of our knowledge 
we are the first to attempt a calculation of $\alpha_{ijk}^{(3)}(k_1,k_2,k_3)$ in the 
squeeezed limit using a background wave method\footnote{Although similar results can be found in \cite{Li:2008gg} using a Hamiltonian method and \cite{Allen:2005ye} using a second-order perturbation theory method. See also \cite{Kehagias:2012pd,Kehagias:2012td} where $\alpha$ in the squeezed limit was given in terms of OPE coefficients. .} which generalizes most easily to a result which is independent of a slow-roll assumption.

We proceed to discuss how to calculate each of these objects now, and then 
combine them to compute the bispectrum of $\zeta$ in the squeezed limit for 
multiple field models.

\subsection{Field evolution between different crossing times} 
\label{evolgen}

The first task is to calculate $\Sigma_{ij}^{(3)}({k_1})$. To do this we need 
to account for the evolution of the 
field-space perturbations on a flat hypersurface with wavenumber $k_1$ between 
the time this wavenumber crosses the horizon and the later time $t_3$.  
We choose to approach this problem in a manner closely connected with the 
$\delta N$ framework, and which therefore provides a unified treatment of the 
overall problem. It will also make transparent the cases in which analytic progress 
can be made.

Returning to the separate universe picture, perturbations at time $t_3$ are defined by
\begin{align}
\delta \phi^{(3)}_{i}(\vect{x}) \equiv  \phi^{(3)}_{i}(\vect{x}) -  \phi^{(3)}_{i}
\end{align}
where $ \phi^{(3)}_{i}(\vect{x})$ is the true value, and $\phi^{(3)}_{i}$ without an $\vect{x}$ argument is the homogeneous background value. In general, for a given 
$i$, $\phi^{(3)}_{i}(\vect{x})$ will depend on the value of all the fields 
and field velocities at the earlier time $t_1$. Assuming, however, slow-roll between 
$t_1$ and $t_3$, the fields velocities become functions of the fields. One 
can write therefore 
\begin{align}
\delta \phi^{(3)}_{i}(\vect{x}) &=  \phi^{(3)}_{i}({\phi}^{(1)}_{j}(\vect{x})) -  \phi^{(3)}_{i} 
\\
&= \phi^{(3)}_{i}(\phi^{(1)}_{j}+ \delta\phi^{(1)}_{j}(\vect{x}) )  -  \phi^{(3)}_{i}.
\end{align}
In analogy with the $\delta N$ expression, one 
can Taylor expand this in the perturbation $\delta\phi^{(1)}_{j}(\vect{x})$ to give
\begin{align}
\delta \phi^{(3)}_{i}(\vect{x}) = \delta\phi^{(1)}_{j}(\vect{x})\frac{\partial \phi^{(3)}_{i}}{\partial \phi^{(1)}_{j}} + ... 
\end{align}
In this paper we won't need the higher order terms. We use the shorthand
\begin{align}
\Gamma^{(3,1)}_{ij}  &\equiv \frac{\partial \phi^{(3)}_{i}}{\partial \phi^{(1)}_{j}}\label{gammaresumm}
\end{align}
which we will often refer to as the `$\Gamma$-matrix'. We would like to emphasise that the `$\Gamma$-matrix' used here is constructed using the background cosmology and is analogous to $N_{i}^{(T)}$ in \eqref{nicoeff}.
In Fourier space we have
\begin{align}
\label{eq:gammaDef}
\delta \phi^{(3)}_{i,\vect{k}} &=  \Gamma^{(3,1)}_{ij} \delta \phi^{(1)}_{j,\vect{k }}  + \dots \, \ \ .
\end{align}
Objects like the $\Gamma$ matrices have been used by a number of authors in the past\footnote{$\Gamma$ can also be generalised to be $k$-dependent, as in \cite{Mulryne:2013uka}.}
\cite{Yokoyama:2007uu,Yokoyama:2007dw,Yokoyama:2008by,Avgoustidis:2011em,2011JCAP...11..005E,Seery:2012vj,Anderson:2012em,Elliston:2012ab}, and we note the properties:  
$\Gamma^{(a,a)}_{ij}=\delta_{ij}$, $\Gamma^{(c,b)}_{ki}\Gamma^{(b,a)}_{ij}=\Gamma^{(c,a)}_{kj}$ and so $\Gamma^{(c,a)}_{ki}\Gamma^{(a,c)}_{ij}=\delta_{kj}$ 
for arbitrary times $t_a$, $t_b$, $t_c$. Moreover 
since $\zeta$ in \eqref{zetaatstar} is independent of $T$, at 
first order we find that 
\begin{align}
N^{(b)}_{i }\delta \phi^{(b)}_{i,\vect{k}} =N^{(a)}_{i }\delta \phi^{(a)}_{i,\vect{k}}
\end{align}
and so we must have
\begin{align}
N^{(b)}_{i } = N^{(a)}_{j }\Gamma^{(a,b)}_{ji} .\label{Ngamma}
\end{align}

We can now use the $\Gamma$ matrices to relate correlation functions at different times, and 
in particular using Eq.~\eref{canonicalsigma} we find

\begin{empheq}[box=\fbox]{align}
\Sigma_{ij}^{(3)}(k_1) &=  \Gamma^{(3,1)}_{ik}\Gamma^{(3,1)}_{jk} \Sigma_{ij}^{(1)}(k_1) 
 =  \Gamma^{(3,1)}_{ik}\Gamma^{(3,1)}_{jk} \frac{{H^{(1)}}^2}{2 k_1^3}\,
\label{sigma331} 
\end{empheq}
While at one level Eq.~\eqref{sigma331} has simply swapped the unknown 
correlation matrix 
$\Sigma_{ij}^{(3)}(k_1)$ for the unknown $\Gamma^{(3,1)}_{ik}$ matrix, the point 
is that the problem has been reduced to 
a matter of solving the background cosmology, which must be done 
anyway to apply the $\delta N$ formalism. Moreover, 
in all models where the derivatives of $N$ can be calculated analytically, 
analytic expressions should also exist for the required $\Gamma$ matrix, as we will see in an explicit 
example below. On the other hand, if numerical 
tools are needed to calculate the derivatives of $N$, similar tools can be applied to calculate 
 $\Gamma^{(3,1)}_{ik}$. The formulation in terms of $\Gamma$ is therefore very convenient  for explicitly studying the squeezed limit.

\subsection{The field-space bispectrum in the squeezed limit}\label{subsec:3ptsqueezed}
\label{subsec:alphacalc}

The next step is to calculate $\alpha_{ijk}^{(3)}(k_1,k_2,k_3)$ in the squeezed limit. To do so we can adapt an idea used by Maldacena \cite{Maldacena:2002vr} and subsequently other authors \cite{Creminelli:2004yq,Cheung:2007sv} in the single field context. The calculation relies on the following simplifying assumption: the long wavelength mode $k_1$ which exits the horizon at the first exit time, $t_1$, can only affect the much shorter wavelength modes, which at that time are still deep inside the horizon, through its effect on the background cosmology. The $k_1$ mode shifts the background field configuration, and hence changes the background cosmology which is felt by the two modes which exit later -- producing a correlation between the $k_1$ mode and the two-point  function of the $k_2,k_3$ modes in the shifted background. 

In the single field case the calculation is performed directly with the modes of 
$\zeta$ in the comoving gauge, and the 
shift in background cosmology induced by the long wavelength mode exiting the 
horizon can be thought of as a \textit{shift in the time} at which 
the short modes exit. This leads to the three-point correlation function of 
$\zeta$ being related to the tilt of the two-point function 
of the short modes, and to the famous consistency relation of Maldacena\footnote{Note that the consistency relation holds for all single field models with a Bunch-Davies initial state and where the classical solution is a dynamical attractor.}.

In the multiple field case we are calculating the three-point function of 
field perturbations in the flat gauge, and the 
shift in the background cosmology induced by the long wavelength mode exiting the 
horizon can be thought of as a \textit{shift in the background field values}. This won't result in a simple consistency relation
for $\zeta$, but does allow analytic progress to be made in calculating $\alpha$ in the highly squeezed limit. 

To calculate $\alpha$ we follow a similar calculation to that in Cheung {\it et al.} \cite{Cheung:2007sv} for $\braket{\zeta\zeta\zeta}$ in the single field case. We show a derivation of the squeezed limit of $\alpha$ which relies on slow-roll - though this derivation can be easily generalised to relax the assumption of slow-roll. In the end the final result we get for the squeezed limit of $\braket{\zeta\zeta\zeta}$ must rely on a slow-roll approximation at the time of the last horizon crossing, $t_3$, 
because we will be using the $\delta N$ formalism at that time.

We begin our calculation in position space, denoting short-wavelength perturbations with a superscript $S$, and long-wavelength perturbations with a superscript $L$. We ask how a 
short-wavelength	 two-point function $\langle \delta {\phi^S}^{(3)}_{j}(\vect{x_2})   \delta {\phi^S}^{(3)}_{k }(\vect{x_3}) \rangle$ at $t_3$ is affected by a further long-wavelength
perturbation $\delta {\phi^L}^{(3)}_{i }(\vect{x})$.
To proceed, we evaluate the long-wavelength fluctuation $\delta {\phi^L}^{(3)}_{i }(\vect{x})$ at the 
midpoint $\vect{x} = \vect{x_+} \equiv (\vect{x_2}+\vect{x_3})/2$. Then we 
Taylor expand the short-wavelength two-point function
around the value it would have in the absence of the 
long-wavelength perturbation, denoting this value 
by a subscript $0$. One finds
\begin{align}
\begin{split}
\langle \delta {\phi^S}^{(3)}_{j}(\vect{x_2})   \delta {\phi^S}^{(3)}_{k }(\vect{x_3}) \rangle\Big|_{\delta {\phi^L}^{(3)}_{m }(\vect{x_+})} =
&\langle \delta {\phi^S}^{(3)}_{j}(\vect{x_2})   \delta {\phi^S}^{(3)}_{k }(\vect{x_3}) \rangle\Big|_0
\\
&+ \delta {\phi^L}^{(3)}_{m}(\vect{x_+})
 \langle \delta {\phi^S}^{(3)}_{j}(\vect{x_2})   \delta {\phi^S}^{(3)}_{k }(\vect{x_3}) \rangle_{,m} \Big|_0 
+ ...
\end{split} \label{Taylorex}
\end{align}
where  the subscript $,m$ denotes the partial derivative with respect to the background field $ \phi^{(3)}_{m }$. At this stage we employ the simple soft limit argument discussed above and assume that in the squeezed limit the three-point function in momentum space will receive its largest contribution from the correlation between the long-wavelength mode, which effectively shifts the background cosmology, and the short-wavelength two-point function in the shifted background. This leads to
\begin{align}
\langle \delta {\phi^L}^{(3)}_{i }(\vect{x_1})\delta {\phi^S}^{(3)}_{j }(\vect{x_2})\delta {\phi^S}^{(3)}_{k }(\vect{x_3}) \rangle 
&\approx \langle \delta {\phi^L}^{(3)}_{i }(\vect{x_1}) \langle \delta {\phi^S}^{(3)}_{j }(\vect{x_2})\delta {\phi^S}^{(3)}_{k }(\vect{x_3}) \rangle\Big|_{\delta {\phi^L}^{(3)}_{m }(\vect{x_+})}  \rangle
\\
&\approx \langle \delta {\phi^L}^{(3)}_{i }(\vect{x_1})\delta {\phi^L}^{(3)}_{m }(\vect{x_+})  \langle \delta {\phi^S}^{(3)}_{j }(\vect{x_2})\delta {\phi^S}^{(3)}_{k}(\vect{x_3}) \rangle_{,m}  \Big|_0 
\rangle
\\
&\approx \langle \delta {\phi^L}^{(3)}_{i }(\vect{x_1})\delta {\phi^L}^{(3)}_{m }(\vect{x_+})\rangle  \langle \delta {\phi^S}^{(3)}_{j }(\vect{x_2})\delta {\phi^S}^{(3)}_{k }(\vect{x_3}) \rangle_{,m}  \Big|_0 
\\
&\approx \int\frac{d^3\vect{p}}{(2\pi)^3}\frac{d^3\vect{q}}{(2\pi)^3}e^{i\vect{p} \cdot(\vect{x_1}-\vect{x_+})+i\vect{q} \cdot(\vect{x_2}-\vect{x_3})}\Sigma_{im}^{(3)}(p)
 \Sigma_{jk,m}^{(3)}(q) \Big|_0\,.
\end{align}
In what follows we will drop the subscript $|_0$ for notational ease. Now we insert $1=\int d^3\vect{k_1}\delta(\vect{k_1} + \vect{p})$ to get
\begin{align}
\begin{split}
&\langle \delta {\phi^L}^{(3)}_{i }(\vect{x_1})\delta {\phi^S}^{(3)}_{j }(\vect{x_2})\delta {\phi^S}^{(3)}_{k }(\vect{x_3}) \rangle 
\\
&\approx \int \frac{d^3\vect{k_1}}{(2\pi)^3}\frac{d^3\vect{p}}{(2\pi)^3}\frac{d^3\vect{q}}{(2\pi)^3} e^{- i\vect{k_1}\cdot \vect{x_1} -i\vect{p} \cdot\vect{x_+}+i\vect{q} \cdot(\vect{x_2}-\vect{x_3})}
 (2\pi)^3\delta(\vect{k_1} + \vect{p}) \Sigma_{im}^{(3)}(p)
 \Sigma_{jk,m}^{(3)}(q).
\end{split}
\end{align}
Changing the 
integration variables from $\vect{p}, \vect{q}$ to $\vect{k_2} =\frac{1}{2}\vect{p}- \vect{q}$ and $\vect{k_3} =\frac{1}{2}\vect{p}+ \vect{q}$ we find
\begin{align}
\begin{split}
&\langle \delta {\phi^L}^{(3)}_{i }(\vect{x_1})\delta {\phi^S}^{(3)}_{j }(\vect{x_2})\delta {\phi^S}^{(3)}_{k }(\vect{x_3}) \rangle 
\\
&\approx \int \frac{d^3\vect{k_1}}{(2\pi)^3} \frac{d^3\vect{k_2}}{(2\pi)^3} \frac{d^3\vect{k_3}}{(2\pi)^3}e^{- i\vect{k_1}\cdot \vect{x_1} - i\vect{k_2}\cdot \vect{x_2}- i\vect{k_3}\cdot \vect{x_3}}
 (2\pi)^3\delta(\vect{k_1} + \vect{k_2}+\vect{k_3})  \Sigma_{im}^{(3)}(k_1)
 \Sigma_{jk,m}^{(3)}(k_3) ,
\end{split} \label{3ptpert}
\end{align}
 where we have used the relation $\vect{q} = \vect{k_3}(1 + {\cal O}(k_1/k_3))$. From \eqref{3ptpert} we can now read off the squeezed limit of the momentum space three-point function of the field perturbations
\begin{align}
\begin{split}
\lim _{k_1 \ll k_3,k_2}\langle \delta \phi^{(3)}_{i, \vect{k_1}}\delta \phi^{(3)}_{j, \vect{k_2}}\delta \phi^{(3)}_{k, \vect{k_3}} \rangle 
&\approx (2\pi)^3\delta(\vect{k_1} + \vect{k_2}+\vect{k_3}) \Sigma_{im}^{(3)}(k_1)
 \Sigma_{jk,m}^{(3)}(k_3)  
\end{split}
\end{align}
and so 
\begin{empheq}[box=\fbox]{align}
\lim _{k_1 \ll k_3,k_2} \alpha_{ijk}^{(3)}(k_1,k_2,k_3) &\approx \Sigma_{im}^{(3)}(k_1)
 \Sigma_{jk,m}^{(3)}(k_3) \label{alphageneral}
\end{empheq}
This is a very general expression for the squeezed limit of $\alpha$, independent of the multiple field model, relating the squeezed limit of the three-point function of the field perturbations to the two-point function of the field perturbations and its derivatives with respect to the background fields. It is one of the principal results of this paper. It only relied on slow-roll at time $t_3$ in \eqref{Taylorex} where the Taylor expansion was done only in terms of the field, rather than both the field and its velocity. If one promoted the indices $i$ to run over both fields and field velocities, then \eqref{alphageneral} with this more general index notation would still hold, independent of whether slow-roll was valid at time $t_3$.

At this stage we can make our result more explicit in the case where slow-roll is valid between 
the horizon exit times $t_1$ and $t_3$. In this case, we can 
use the $\Gamma$ evolution between crossing times, which by \eqref{sigma331} and \eqref{canonicalsigma} gives
\begin{align}
\begin{split}
\alpha_{ijk}^{(3)}(k_1,k_2,k_3)
&\approx \Gamma^{(3,1)}_{il}\Gamma^{(3,1)}_{mn} \Sigma_{ln}^{(1)}(k_1)
 \Sigma_{jk,m}^{(3)}(k_3) 
 \approx \Gamma^{(3,1)}_{il}\Gamma^{(3,1)}_{ml}\delta_{jk}\frac{{H^{(1)}}^2}{2k_1^3}    \frac{\left [{{H^{(3)}}}^2\right ]_{,m}}{2k_3^3}
 \\ 
 &\approx -\Gamma^{(3,1)}_{il}\Gamma^{(3,1)}_{ml}\delta_{jk}\frac{{H^{(1)}}^2}{2k_1^3}    \frac{ {{H^{(3)}}}^2}{2k_3^3}\frac{d \phi^{(3)}_m}{d N}. 
 \end{split}\label{squeezealpha}
\end{align}
where the final line uses the slow-roll equations of motion for the background fields, and for clarity we note that $dN \equiv d \log a$ is the local measure of e-folding time (and is not related to the $\delta N$ formula).

\subsection{The squeezed limit of the bispectrum of $\zeta$}

\begin{figure}[h!]
\center
\includegraphics[scale=0.45]{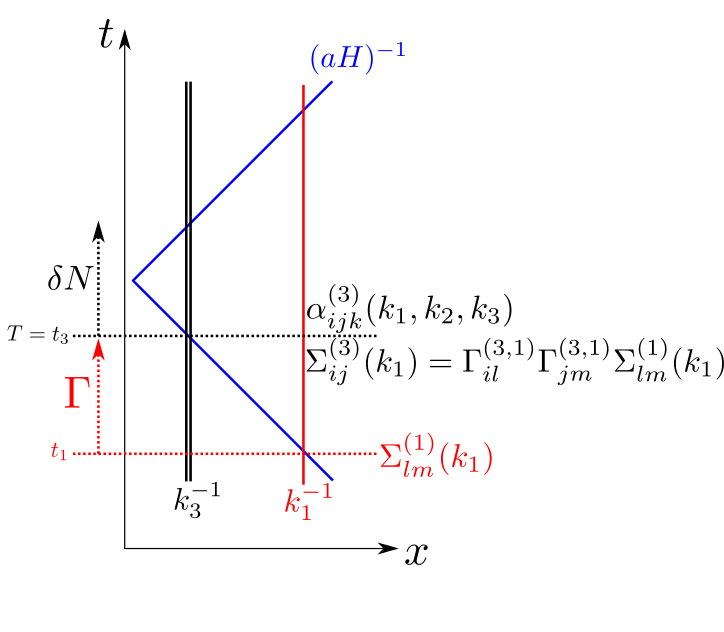} 
\caption{Schematic picture of $\delta N$ evolution from time $T=t_3$ onwards, using $\Sigma^{(3)}_{ij}(k_1)$ from Subsection~\ref{evolgen} and $\alpha_{ijk}^{(3)}(k_1,k_2,k_3)$ from Subsection~\ref{subsec:alphacalc}. The $\Gamma$-evolution of field perturbations occurs between $t_1$ and $t_3$. The blue line is the comoving Hubble radius, the solid red line is the inverse of the squeezed wavenumber, $k_1$, while the solid black lines are the inverses of the other wavenumbers, $k_2$ and $k_3$.}
\label{fig:schem}
\end{figure}

We now have all the ingredients we need to calculate the bispectrum of $\zeta$  
\eqref{bispectrumdeltaN} in the squeezed limit for multiple field inflation, and since we now use the $\delta N$ formalism, the results which follow in this and subsequent sections are only valid in slow-roll models. 
The schematic picture of our approach is shown in Figure~\ref{fig:schem}.

Putting $\alpha_{ijk}^{(3)}(k_1,k_2,k_3)$ from \eqref{squeezealpha} and $\Sigma^{(3)}_{ij}(k_1)$ from \eqref{sigma331} into \eqref{bispectrumdeltaN}, with $T\mapsto t_3$, gives 
\begin{align}
\begin{split}
\lim _{k_1 \ll k_2,k_3} B_{\zeta}(k_1,k_2,k_3) 
\approx \ &-N^{(3)}_{i}N^{(3)}_{j}N^{(3)}_{j} \Gamma^{(3,1)}_{ik}\Gamma^{(3,1)}_{mk}\frac{{H^{(1)}}^2}{2k_1^3}   \frac{ {{H^{(3)}}}^2}{2k_3^3}\frac{d \phi^{(3)}_m}{d N}
\\
&+N^{(3)}_{i}N^{(3)}_{jk}N^{(3)}_{k}\left [2 \Gamma^{(3,1)}_{im}\Gamma^{(3,1)}_{jm}\frac{{H^{(1)}}^2}{2k_1^3}\frac{{{H^{(3)}}}^2}{2k_3^3} + \delta^{ij} \left ( \frac{{{H^{(3)}}}^2}{2k_3^3}\right)^2         \right   ]
\end{split} \label{bispectrumsqueezed}
\end{align}
and since $k_1 \ll k_3$ this can be simplified to
\begin{empheq}[box=\fbox]{align}
\begin{split}
\lim _{k_1 \ll k_2,k_3} B_{\zeta}(k_1,k_2,k_3) 
\approx \ &-N^{(3)}_{i}N^{(3)}_{j}N^{(3)}_{j} \Gamma^{(3,1)}_{ik}\Gamma^{(3,1)}_{mk}\frac{{H^{(1)}}^2}{2k_1^3}   \frac{ {{H^{(3)}}}^2}{2k_3^3}\frac{d \phi^{(3)}_m}{d N}
\\
&+2N^{(3)}_{i}N^{(3)}_{jk}N^{(3)}_{k}  \Gamma^{(3,1)}_{im}\Gamma^{(3,1)}_{jm}\frac{{H^{(1)}}^2}{2k_1^3}\frac{{{H^{(3)}}}^2}{2k_3^3} \label{bispectrumsqueezed2}
\end{split}
\end{empheq}
which is one of the main results of this paper. In Appendix~\ref{App:SF} we check that it reduces to the Maldacena result \cite{Maldacena:2002vr} in the single field limit. In Appendix~\ref{App:SL} we check that our result for $\alpha$ in \eqref{squeezealpha} agrees with the result of Seery \& Lidsey  \cite{Seery:2005gb} if we take a near-equilateral limit. 

We can then form the reduced bispectrum in the squeezed limit
\begin{empheq}[box=\fbox]{align}
\begin{split}
&\lim _{k_1 \ll k_2,k_3} \frac{6}{5}\fnl(k_1,k_2,k_3) 
\approx-\frac{N^{(3)}_{i} \Gamma^{(3,1)}_{ik}\Gamma^{(3,1)}_{mk} }
{2N^{(3)}_{l}N^{(3)}_{n}
 \Gamma^{(3,1)}_{lq}\Gamma^{(3,1)}_{nq} 
} \frac{d \phi^{(3)}_m}{d N}
 +\frac{N^{(3)}_{i}N^{(3)}_{jk}N^{(3)}_{k}   \Gamma^{(3,1)}_{im}\Gamma^{(3,1)}_{jm}  }
{N^{(3)}_{l}N^{(3)}_{n}N^{(3)}_{p}N^{(3)}_{p}
   \Gamma^{(3,1)}_{lq}\Gamma^{(3,1)}_{nq}        
} .
\end{split} \label{fnlsqueezedorig}
\end{empheq}

For convenience we define 
\begin{align}
B^{[\alpha]}(k_1,k_2,k_3) &\equiv -N^{(3)}_{i}N^{(3)}_{j}N^{(3)}_{j} \Gamma^{(3,1)}_{ik}\Gamma^{(3,1)}_{mk}\frac{{H^{(1)}}^2}{2k_1^3}   \frac{ {{H^{(3)}}}^2}{2k_3^3}\frac{d \phi^{(3)}_m}{d N}
\\
\frac{6}{5
}\fnl^{[\alpha]}(k_1,k_2,k_3) &\equiv -\frac{N^{(3)}_{i} \Gamma^{(3,1)}_{ik}\Gamma^{(3,1)}_{mk} }
{2N^{(3)}_{l}N^{(3)}_{n}
 \Gamma^{(3,1)}_{lq}\Gamma^{(3,1)}_{nq} 
} \frac{d \phi^{(3)}_m}{d N}
\end{align}
so that superscript $[\alpha]$ labels the contribution from the term involving $\alpha$.

As we noted in \S\ref{subsec:deltaN} for the near-equilateral configuration, the term coming from $\alpha$ can only be of order slow-roll \cite{Seery:2005gb}, and so if the reduced bispectrum of $\zeta$ is to be sufficiently large to be observable by present or next generation experiments, it will be dominated by the second line of Eq.~(\ref{bispectrumdeltaN}) \cite{Lyth:2005qj}.

For our highly squeezed case it's not immediately clear whether the contribution of $\fnl^{[\alpha]}(k_1,k_2,k_3)$ must be of order 
$\epsilon$. One might think that because at least one component of the vector 
$N_i$ is of order $\epsilon^{-1/2}$, and because as many $\Gamma $ matrices appear 
in the numerator as in the denominator, that the likely order is indeed $\epsilon$. However, while this is likely the most common outcome, because $\Gamma^{(3,1)}_{ij}$ is a matrix and appears with different contractions in the numerator compared with the demoninator, it is not impossible that the contractions may conspire to make $\fnl^{[\alpha]}(k_1,k_2,k_3)$ larger than $\mathcal{O}(\epsilon)$. 
In this paper we do not perform a general study, but armed with this explicit expression for $\fnl^{[\alpha]}(k_1,k_2,k_3)$ we 
can consider the amplitude of this term on a case by case basis.

Finally, for completeness, we note that in addition to purely scalar correlations, it is possible to calculate the three-point functions involving both the scalar curvature perturbation and gravitons in the squeezed limit. In Appendix~\ref{app:grav} we use similar techniques to those employed above to find the scalar-graviton three-point functions for multiple field models, noting that the graviton-only three-point function will be the same as for the single field case, as given in \cite{Maldacena:2002vr}.

\section{Scale dependence}\label{sec:scaledep}

\subsection{Spectral index of the halo bias} \label{sec:halo}
An important quantity for large scale structure surveys is the \textit{scale-dependent halo bias}, $\delta b(k_1)$ \cite{PhysRevD.77.123514}, which is sensitive to how the ratio of the bispectrum to the power spectrum, $B_{\zeta}(k_1,k_2,k_3)/P_{\zeta}(k_1)$, scales with the squeezed momentum $k_1$, where $k_1 \ll k_2\approx k_3$. This is captured by the \textit{spectral index of the halo bias}, $n_{\delta b} \equiv n_{\text{sq}} - (n_s-1)$, where $n_{\text{sq}}$ is the tilt
of the squeezed limit of the bispectrum with respect to its squeezed momentum $k_1$:
\begin{align}
\lim _{k_1 \ll k_2,k_3} B_{\zeta}(k_1,k_2,k_3) \sim \frac{\mathcal{B}_\zeta}{k_1^3 k_3^3}\left (\frac{k_1}{k_s}\right )^{n_{\text{sq}}}
\end{align}
with $k_s$ some arbitrary scale and $\mathcal{B}_\zeta$ roughly constant. Dias {\t et al.} \cite{Dias:2013rla} investigated $n_{\delta b}$ in multiple field inflation in the case where $|\log(k_1/k_3)|$ is 
of order a few, as discussed further in Appendix~\ref{App:DRSreduction}. Here we would like 
to explore the highly squeezed case, and find if there are significant differences.

Our results of \S\ref{sec:deltaNdiffcross} for the bispectrum in the highly squeezed limit can be applied to calculate $n_{\delta b}$ for large values of $|\log(k_1/k_3)|$. Differentiating $\log (k_1^3 B_{\zeta})$ in \eqref{bispectrumsqueezed2} with respect to $\log k_1$ we find
\begin{align}
n_{\text{sq}} = -\frac{\left [N^{(3)}_{i}N^{(3)}_{q}(  N^{(3)}_{q}V^{(3)}_{,j} + 6N^{(3)}_{qj}{{H^{(3)}}}^2    )\right ]}{\left [N^{(3)}_{m}N^{(3)}_{r}(  N^{(3)}_{r}V^{(3)}_{,n} + 6N^{(3)}_{rn}{{H^{(3)}}}^2    )\right ]}
\frac{ \left [2\epsilon^{(1)}L^{(3,1)}_{ij} -  P^{(3,1)}_{ij,1}  \right ]}{ L^{(3,1)}_{mn}} \label{nsq}
\end{align}
where we have defined
\begin{align}
L^{(3,1)}_{ij}&\equiv \Gamma^{(3,1)}_{im}\Gamma^{(3,1)}_{jm}\label{lij}
\\
P^{(3,1)}_{ij,1} &\equiv \frac{d  L^{(3,1)}_{ij}}{d \log k_1} = - \frac{V^{(1)}_{,l}}{{V^{(1)}}}\left (\Gamma^{(3,1)}_{ik,l}\Gamma^{(3,1)}_{jk} +\Gamma^{(3,1)}_{ik}\Gamma^{(3,1)}_{jk,l}\right ) \label{pij1}
\\
\Gamma^{(3,1)}_{ik,l} &\equiv \frac{\partial}{\partial \phi^{(1)}_l}\Gamma^{(3,1)}_{ik}\label{gammaikl}
\end{align}
and we have used the slow-roll equations to relate $\dot{\phi}_i$ to $V_{,i}$ in \eqref{pij1}.

The spectral index of the halo bias requires also the tilt of the power spectrum at $k_1$
\begin{align}
n_s -1&= \frac{d \log (k_1^3 P_{\zeta}(k_1))}{d \log k_1}
= -2\frac{N^{(3)}_{i}N^{(3)}_{j} \Gamma^{(3,1)}_{ik}\Gamma^{(3,1)}_{jl}M^{(1)}_{kl} }{N^{(3)}_{m}N^{(3)}_{n}L^{(3,1)}_{mp}} \label{ns}
\end{align}
where
\begin{align}
M^{(1)}_{ij} &\equiv \epsilon^{(1)} \delta_{ij} + u^{(1)}_{ij} \label{Mdef}
\\
\text{ and } u^{(1)}_{ij} &\equiv \frac{V^{(1)}_{,i}V^{(1)}_{,j}}{{V^{(1)}}^2} - \frac{V^{(1)}_{,ij}}{V^{(1)}}. \label{udef}
\end{align}
This leads to the expression for the spectral index of the halo bias
\begin{empheq}[box=\fbox]{align}
\begin{split}
n_{\delta b} = &-\frac{\left [N^{(3)}_{i}N^{(3)}_{q}(  N^{(3)}_{q}V^{(3)}_{,j} + 6N^{(3)}_{qj}{{H^{(3)}}}^2    )\right ]}{\left [N^{(3)}_{m}N^{(3)}_{r}(  N^{(3)}_{r}V^{(3)}_{,n} + 6N^{(3)}_{rn}{{H^{(3)}}}^2    )\right ]}
\frac{ \left [2\epsilon^{(1)}L^{(3,1)}_{ij} -  P^{(3,1)}_{ij,1}  \right ]}{ L^{(3,1)}_{mn}}
\\
&+2\frac{N^{(3)}_{i}N^{(3)}_{j} \Gamma^{(3,1)}_{ik}\Gamma^{(3,1)}_{jl}M^{(1)}_{kl} }{N^{(3)}_{m}N^{(3)}_{n}L^{(3,1)}_{mn}} 
\end{split}\label{eq:nsqgamma}
\end{empheq}
valid for large values of $|\log(k_1/k_3)|$, and is another key result of this paper. 
In Appendix~\ref{App:DRSreduction} we show that when we take our exit times to be roughly equal \eqref{eq:nsqgamma} reduces to the same form of (17) of \cite{Dias:2013rla}, given in the appendix as \eqref{DRSnsq}, and in \S \ref{multiplesource} we compare approaches for a concrete model.

We note that if $\fnl^{[\alpha]}(k_1,k_2,k_3)$ is negligible, then our expression simplifies to 
\begin{align}
n_{\delta b} = -\frac{N^{(3)}_{i}N^{(3)}_{q} N^{(3)}_{qj}   }{N^{(3)}_{m}N^{(3)}_{r}N^{(3)}_{rn}  }
\frac{ \left (2\epsilon^{(1)}L^{(3,1)}_{ij} -  P^{(3,1)}_{ij,1}  \right )}{ L^{(3,1)}_{mn}}
+2\frac{N^{(3)}_{i}N^{(3)}_{j} \Gamma^{(3,1)}_{ik}\Gamma^{(3,1)}_{jl}M^{(1)}_{kl} }{N^{(3)}_{m}N^{(3)}_{n}L^{(3,1)}_{mp}} .\label{ndbsimple}
\end{align}

\subsection{Tilts of the reduced bispectrum in the squeezed limit}\label{sec:tilt}
In a similar manner to Refs.~\cite{2010JCAP...02..034B,Byrnes:2008zy,Sefusatti:2009xu,Byrnes:2009qy}, one can study the scaling of the squeezed limit of the reduced bispectrum \eqref{fnlsqueezedorig} with respect to $k_1$ and $k_3$. In the squeezed limit $k_2\approx k_3$, 
so one can parametrize how this depends on scale by differentiating with respect to $k_1$ or $k_3$ leading to the tilts 
\begin{align}
n_{\fnl}^{X} \equiv \frac{\partial \log |\fnl(k_1,k_2,k_3)|}{ \partial \log X}
\end{align}
where $X=k_1$ or $k_3$, with the other momenta held fixed in the derivative. 
Note that $n_{\fnl}^{k_1}=n_{\delta b}$ since the $k_1$ dependence of $\fnl$ is captured by the scaling of the ratio $B_{\zeta}(k_1,k_2,k_3)/P_{\zeta}(k_1)$ with $k_1$.
We calculate $n_{\fnl}^{k_3}$ in \eqref{nfnlk3} of Appendix~\ref{app:tilt}.
We note that each of these results are different to the scale dependence of the 
equilateral configuration of Byrnes {\it et al.} \cite{2010JCAP...02..034B} given in Eq.~\eqref{nfnleq}, since we are working only in the squeezed limit. 
These authors also considered the scale dependence of near-equilateral triangles, writing the wavenumbers as $k_a=\alpha_a \tilde{k}$, and varying with respect to $\tilde{k}$, keeping the $\alpha_a$ constant.
Their near-equilateral result has the same form as \eqref{nfnleq}, with all $*$'s replaced by the exit time of a pivot scale $k_p$ not too different from the $k$'s. Their calculation relies on an expansion to first order in $|\log (k_1/k_3)|$ and as a result is only valid for a small squeezing. 
We calculate the $\tilde{k}$ tilt using our approach in Eq~\eref{nfnltildek} of Appendix~\ref{app:tilt}, where more details on the expansion of \cite{2010JCAP...02..034B} can also be found. We emphasise again that all of our expressions can be employed for a large hierarchy of scales.

\section{Employing the $\Gamma$ formalism in concrete models}\label{sec:gammaimport}

In this paper we have advocated the use of the $\Gamma$ matrices to allow 
the bispectrum of $\zeta$ to be calculated in the highly squeezed limit for multiple 
field models. These 
matrices allow us to account for the evolution of the inflationary fields 
between horizon crossing times, and 
help provide compact expressions for the bispectrum and its scale dependence, including the 
contribution from the field-space three-point function. To be of use, however, we must be
able to calculate the $\Gamma$ matrices in concrete settings.

In any given model, we could solve for $\Gamma^{(3, 1)}_{ij}$ numerically. Either by 
solving the equation of motion it satisfies \cite{2011JCAP...11..005E} from the initial conditions at 
$t_1$ of $\Gamma^{(3,1)}_{ij}= \delta_{ij}$, which follow 
from its definition, or by solving the background equations and implying finite differences as was 
done for the derivatives of $N$ in, for example, Refs.~\cite{Vernizzi:2006ve,2012JCAP...09..008L,2011JCAP...11..005E}. However, what 
makes this parametrisation particularly useful is that in any model for which the 
derivatives of $N$ can be calculated analytically, the $\Gamma$ matrices also admit 
analytic solutions. This allows us to compute  the amplitude of the highly squeezed limit of 
the bispectrum in such models and to compare this against previous squeezed limit expressions, as well as against the amplitude of the bispectrum for near-equilateral triangles. 

We therefore calculate 
$\Gamma$ analytically in \S\ref{sumsep} for sum-separable potentials. Then we consider the importance 
of $\Gamma$ in specific settings. 
In \S\ref{singlesource} we show that in all single-source models, where only one field (which need not be the inflaton) contributes towards the curvature perturbation, 
the effect of the $\Gamma$ matrices is to cause the reduced bispectrum in the squeezed 
limit to become independent of the squeezed momentum $k_1$, and that because the bispectrum scales with $k_1$ in exactly the same way as the power spectrum, the spectral index of the halo bias will be zero, and so not  observable. Then in 
\S\ref{multiplesource} we consider a specific multiple-source model where more interesting 
results are possible. In particular, we examine a mixed curvaton-inflaton 
model \cite{Lyth:2001nq,Lyth:2002my,Langlois:2004nn,Sasaki:2006kq,Fonseca:2011aa,Fonseca:2012cj,Meyers:2013gua,Elliston:2014zea,Byrnes:2015asa} allowing for self-interaction terms for the curvaton. For this specific model with our given parameter choices, 
we find that in highly squeezed cases relevant for future observations, the bispectrum is suppressed by the $\Gamma$ matrices at a level of 20\% when compared to using the 
existing expressions of Byrnes {\it et al.} \cite{2010JCAP...02..034B} and Dias {\it et al.} \cite{Dias:2013rla} for this model. In addition, we find that the spectral index of the 
halo bias is enhanced at a level of 20\% in this model compared with the results that would be obtained with previous expressions which assume a mild hierarchy of scales.

\subsection{Calculating $\Gamma$: sum-separable potential} \label{sumsep}

In the $\delta N$ framework, all models which are analytically tractable have a common feature. This is
 that the inflationary potentials are of separable form, either sum or product separable \cite{Battefeld:2006sz,Choi:2007su,Vernizzi:2006ve}, or 
  of the  generalised 
 sum-separable form of Ref~\cite{2010PhRvD..82l3515W}. This is true not only for models 
 in which the evolution is tracked during 
 inflation, but also for models in which the post inflationary evolution is important such as the 
 curvaton model.   
 In this work we will focus on sum-separable potentials and 
confirm that we can derive analytic formulae for the $\Gamma$ matrices, using similar techniques as those 
originally 
used for derivatives of $N$ in Ref.~\cite{Vernizzi:2006ve}. 

We will initially work with the simple case of a two field model: $\phi$, $\chi$ and 
write the potential as $W(\phi,\chi)=U(\phi) + V(\chi)$. 
The slow-roll equations are then
\begin{align}
3H \dot{\phi} = -U_{,\phi} \ , \qquad 3H \dot{\chi} = -V_{,\chi} \ , \qquad 3H^2 = W.
\end{align}
Using these slow-roll equations we have
\begin{align}
\frac{ d{\phi}}{ d{\chi}} = \frac{\dot{\phi}}{\dot{\chi}} = \frac{U_{,\phi}}{V_{,\chi} } \label{dphidchi}\,.
\end{align}
The number of e-folds between a flat hypersurface at time $t_1$ and another flat hypersurface at time $t_3$ is 
\begin{align}
\Delta N \equiv \int_{t_1}^{t_3} H dt = \int_{\phi^{(1)}}^{\phi^{(3)}} \frac{H}{\dot{\phi}} d\phi = -\int_{\phi^{(1)}}^{\phi^{(3)}} \frac{W}{U_{,\phi}} d\phi
\end{align}
and using \eqref{dphidchi} this gives
\begin{align}
\Delta N = -\int_{\phi^{(1)}}^{\phi^{(3)}} \frac{U}{U_{,\phi}} d\phi-\int_{\chi^{(1)}}^{\chi^{(3)}} \frac{V}{V_{,\chi}} d\chi. \label{DeltaN13}
\end{align}
We also have, by \eqref{dphidchi}, that
\begin{align}
\int_{\phi^{(1)}}^{\phi^{(3)}} \frac{1}{U_{,\phi}} d\phi = \int_{\chi^{(1)}}^{\chi^{(3)}} \frac{1}{V_{,\chi}} d\chi. \label{UVrel}
\end{align}
To determine $\Gamma^{(3,1)}_{ij}$ we need to find the following four derivatives of flat hypersurface fields
\begin{align}
\Gamma^{(3,1)}_{ij}= \left (\begin{array}{ll}
\frac{\partial \phi^{(3)}}{\partial \phi^{(1)}} & \frac{\partial \phi^{(3)}}{\partial \chi^{(1)}}
\\
\frac{\partial \chi^{(3)}}{\partial \phi^{(1)}} & \frac{\partial \chi^{(3)}}{\partial \chi^{(1)}}
\end{array} \right ).
\end{align}

For flat hypersurfaces, by definition, if we vary our position on the initial slice, then $\Delta N$ 
does not alter. This implies that the derivative 
of $\Delta N$ with respect to field values on the initial flat hypersurface satisfies: $\Delta N _{,\phi ^{(1)} } = 0 = \Delta N _{,\chi^{(1)}}$.  Employing \eqref{DeltaN13} then leads to two independent equations relating the derivatives. 
Moreover, differentiating \eqref{UVrel} with respect to the field values on the initial flat 
hypersurface  
yields two further independent equations relating the derivatives. Between
these four equations we can then solve for each derivative and hence 
determine $\Gamma^{(3,1)}_{ij}$.

We begin with $\Delta N _{,\phi^{(1)}} = 0$ and $\Delta N _{,\chi^{(1)}} = 0$ giving respectively
\begin{align}
-\frac{U^{(3)}}{U^{(3)}_{,\phi}}\frac{\partial \phi^{(3)}}{\partial \phi^{(1)}} - \frac{{V^{(3)}}}{V^{(3)}_{,\chi}}\frac{\partial \chi^{(3)}}{\partial \phi^{(1)}} &= -\frac{U^{(1)}}{U^{(1)}_{,\phi}} \label{dnphi}
\\
-\frac{U^{(3)}}{U^{(3)}_{,\phi}}\frac{\partial \phi^{(3)}}{\partial \chi^{(1)}} - \frac{{V^{(3)}}}{V^{(3)}_{,\chi}}\frac{\partial \chi^{(3)}}{\partial \chi^{(1)}} &= -\frac{{V^{(1)}}}{V^{(1)}_{,\chi}}.\label{dnchi}
\end{align}
Next differentiating \eqref{UVrel} with respect to $\phi^{(1)}$ and $\chi^{(1)}$ gives respectively
\begin{align}
\frac{1}{U^{(3)}_{,\phi}}\frac{\partial \phi^{(3)}}{\partial \phi^{(1)}} - \frac{1}{V^{(3)}_{,\chi}}\frac{\partial \chi^{(3)}}{\partial \phi^{(1)}} &= \frac{1}{U^{(1)}_{,\phi}}\label{dcphi}
\\
\frac{1}{U^{(3)}_{,\phi}}\frac{\partial \phi^{(3)}}{\partial \chi^{(1)}} - \frac{1}{V^{(3)}_{,\chi}}\frac{\partial \chi^{(3)}}{\partial \chi^{(1)}} &= -\frac{1}{V^{(1)}_{,\chi}}.\label{dcchi}
\end{align}
Solving, we find
\begin{align}
\Gamma^{(3,1)}_{ij}= \left (\begin{array}{ll}
\frac{\partial \phi^{(3)}}{\partial \phi^{(1)}} & \frac{\partial \phi^{(3)}}{\partial \chi^{(1)}}
\\
\frac{\partial \chi^{(3)}}{\partial \phi^{(1)}} & \frac{\partial \chi^{(3)}}{\partial \chi^{(1)}}
\end{array} \right ) 
= \left (\begin{array}{ll}
\frac{U^{(3)}_{,\phi}}{U^{(1)}_{,\phi}}\frac{(U^{(1)}+V^{(3)})}{W^{(3)}}\qquad   
& \frac{U^{(3)}_{,\phi}}{V^{(1)}_{,\chi}}\frac{({V^{(1)}}-V^{(3)})}{W^{(3)}}
\\
\frac{V^{(3)}_{,\chi}}{U^{(1)}_{,\phi}}\frac{(U^{(1)}-U^{(3)})}{W^{(3)}}\qquad   
& \frac{V^{(3)}_{,\chi}}{V^{(1)}_{,\chi}}\frac{({V^{(1)}}+U^{(3)})}{W^{(3)}}
\end{array} \right )\,, \label{gammasep}
\end{align}
which is the analytic calculation of $\Gamma$ for two-field sum-separable potentials we required. Note that $i$ labels the rows and $j$ labels the columns.

For a model with $n$ fields one can calculate the $\Gamma$ matrix by generalizing the two-field case. In what follows we will suspend the summation convention of repeated indices, and where a sum should be taken we explicitly state this. We take the sum-separable potential $W(\phi_1,\phi_2,...,\phi_n) = \sum_l W_l(\phi_l)$. We have, in analogy with \eqref{DeltaN13},
\begin{align}
\Delta N = -\sum_l \int_{\phi_l^{(1)}}^{\phi_l^{(3)}} \frac{W_l}{W_{l}'} d\phi_l
\end{align}
where $W_{l}' \equiv \partial W_{l}/ \partial \phi_l$
for which we can write the equations $\Delta N _{,\phi_j^{(1)} } = 0$ in analogy to \eqref{dnphi} and \eqref{dnchi}. 
We also have the relations, in analogy to \eqref{UVrel}, for all $l, i$
\begin{align}
\int_{\phi_l^{(1)}}^{\phi_l^{(3)}} \frac{d\phi_l}{W_{,\phi_l}}  = \int_{\phi_i^{(1)}}^{\phi_i^{(3)}} \frac{d\phi_i}{W_{i,\phi_i}} 
\end{align}
for which we can take derivatives with respect to $\phi_j^{(1)}$ giving equations analogous to \eqref{dcphi} and \eqref{dcchi}. Combining these equations with those from $\Delta N _{,\phi_j^{(1)} } = 0$, after some algebra we arrive at
\begin{align}
\Gamma^{(3,1)}_{ij}=\frac{\partial \phi_i^{(3)}}{\partial \phi_j^{(1)}} =
\frac{W_i'^{(3)}}{W_j'^{(1)}} \frac{(W_j^{(1)}-W_j^{(3)})}{W^{(3)}} + \delta_{ij}\frac{W_i'^{(3)}}{W_i'^{(1)}}
\end{align}
which reduces to \eqref{gammasep} in the two-field case.

\subsection{Single-source models} \label{singlesource}
In what follows we will assume for now that in the highly squeezed limit we can neglect $B_{\zeta}^{[\alpha]}(k_1,k_2,k_3)$, the 
contribution to the bispectrum from intrinsic field-space three-point function. This needs to be checked on a case by case basis.

In the single-source case where only one field, which we denote $\chi$, contributes to $\zeta$, the bispectrum \eqref{bispectrumsqueezed2} (without $B_{\zeta}^{[\alpha]}(k_1,k_2,k_3)$) simplifies to 
\begin{align}
\begin{split}
\lim _{k_1 \ll k_2,k_3} B_{\zeta}(k_1,k_2,k_3) 
\approx &2N^{(3)}_{\chi}N^{(3)}_{\chi\chi}N^{(3)}_{\chi}  \Gamma^{(3,1)}_{\chi m}\Gamma^{(3,1)}_{\chi m}\frac{{H^{(1)}}^2}{2k_1^3}\frac{{{H^{(3)}}}^2}{2k_3^3}. \label{bispectrumsqueezedss}
\end{split}
\end{align}
Now we can use \eqref{Ngamma} to relate $N^{(3)}_{\chi} $ to $N^{(1)}_{i}$ through the $\Gamma$'s to give the bispectrum and reduced bispectrum as
\begin{align}
\lim _{k_1 \ll k_2,k_3} B_{\zeta}(k_1,k_2,k_3) 
&\approx 2N^{(1)}_{m}N^{(3)}_{\chi\chi}N^{(1)}_{m} \frac{{H^{(1)}}^2}{2k_1^3}\frac{{{H^{(3)}}}^2}{2k_3^3} \label{bispectrumsqueezedss2}
\\
&\approx 2P_{\zeta}(k_1)P_{\zeta}(k_3)\dfrac{N^{(3)}_{\chi\chi}}{{N^{(3)}_{\chi}}^2}
\\
\lim _{k_1 \ll k_2,k_3} \frac{6}{5}f_{\text{NL}}(k_1,k_2,k_3) 
&\approx\dfrac{N^{(3)}_{\chi\chi}}{{N^{(3)}_{\chi}}^2}.\label{eqsinglesource}
\end{align}
We note that the effect of the $\Gamma$ matrices has been to make the bispectrum proportional to $2P_{\zeta}(k_1)P_{\zeta}(k_3)$, so that the reduced bispectrum is independent of 
$k_1$, and coincidentally of the same form as would be derived in the near-equilateral regime
assuming $t_* \approx t_3$. Because the bispectrum scales with $k_1$ in exactly the same way as the power spectrum, the spectral index of the halo bias will be zero, and so will not be observable unless 
the intrinsic contribution we have neglected is important. 

Note that the results of this subsection did not rely on assuming a sum-separable potential -- all single source models satisfy \eqref{eqsinglesource}. In the next subsection we will look at a specific multiple-source model which does rely on the assumption of a sum-separable potential.

\subsection{Multiple-source models: the mixed curvaton-inflaton model} \label{multiplesource}

We now consider multiple-source models, and will consider the concrete example of 
a curvaton model for which both the inflaton, $\phi$, and curvaton, $\chi$, contribute towards $\zeta$.
In order to make analytic progress we will 
need expressions for the derivatives of $N$ in this multiple-source model to 
combine with our analytic expression for $\Gamma$, valid for sum-separable potentials. 

More detailed studies of the curvaton scenario can be found in the literature, see e.g. \cite{Lyth:2001nq,Lyth:2002my,Langlois:2004nn,Sasaki:2006kq,Fonseca:2011aa,Fonseca:2012cj}. Here we just give a very brief account outlining our parameter choices and quote the results for the $N$ derivatives. Our main focus will instead be on the differences between our new expressions compared to existing formulae in the literature. We will show these diffences graphically, where the plots are produced using the expressions for the N derivatives that follow, together with the $\Gamma$ matrix for sum-separable potentials.

For the inflaton and curvaton, we take the potential
\begin{align}
W(\phi,\chi) &= \frac{1}{2}m_{\phi}^2 \phi^2 + \frac{1}{2}m_{\chi}^2 \chi^2  + \lambda\chi^n,\label{quadpot}
\end{align}
where the curvaton is given a self-interaction with coupling $\lambda$, and $n>2$.
If we are to see a difference between the bispectrum in the highly squeezed limit 
and the near-equilateral limit, it is natural to expect that we will need the field configuration to differ significantly between different exit times, otherwise the $\delta\phi_{i}$ would not evolve between $t_1$ and $t_3$ and we would just get $\Gamma_{ij}^{(3,1)} = \delta_{ij}$. This means that we expect to see interesting effects for models which generate a significantly scale-dependent non-Gaussianity. In the present model, this means that the curvaton field, while light, cannot be completely frozen. To achieve this we take the mass of the curvaton and inflaton to be the same $m_{\chi}= m_{\phi}$. 

The self-interacting curvaton model can give significant scale dependence of $\fnl$,  \cite{Byrnes:2011gh, Byrnes:2015asa}, and analytic expressions for the derivatives of $N$ are available 
in the limit of weak self-interaction $s_k \equiv 2\lambda\chi_k^{n-2}/m_{\chi}^2 \ll 1$ \cite{Byrnes:2015asa}. Here, and in what follows, a subscript $k$ indicates the result is a function of the value of the curvaton and/or inflaton fields at the time, $t_k$, when a given $k$-mode exits. 

The relevant $N$ derivatives are given by \cite{Lyth2005,2012JCAP...06..028F,Byrnes:2015asa}
\begin{align}
N_{\phi} & = \frac{1}{\sqrt{2\epsilon_\phi}|_k}
\\
N_{\chi} &= \frac{2 r_{\text{dec}}}{3}\frac{\sigma_{\mathrm{osc}}'}{\sigma_{\mathrm{osc}}}\Big|_k
\\
N_{\chi\chi} &= \frac{r_{\text{dec}}}{3}\left [\frac{\sigma_{\mathrm{osc}}''}{\sigma_{\mathrm{osc}}}+ \left ( \frac{\sigma_{\mathrm{osc}}'}{\sigma_{\mathrm{osc}}}\right )^2\right ]\Big|_k
\\
\text{where }  \sigma_{\mathrm{osc}}(\chi_k) &\propto \chi_k\left (1 + \frac{n}{2}s_k \right )^{-1/(n-2)}.
\end{align}
The parameter $r_{\text{dec}}$ denotes the value of $3\rho_{\chi}/(3\rho_{\chi} + 4 \rho_{\gamma})$ at the time of curvaton decay, where $\rho$ is the energy density and the subscript labels the species, with $\gamma$ denoting radiation. In the example which follows 
we take to be $r_{\text{dec}}=0.02$. We have neglected $N_{\phi\phi}, N_{\phi\chi}$ which are much smaller than $N_{\chi\chi}$ for the parameter choices considered.

We take the initial condition\footnote{Note we are still working in units where the reduced Planck mass is set to one.} $\phi_0 = 16$ which leads to $63.5$ 
e-folds of inflation and assume that all scales which exit the horizon after this time 
are within the horizon today and potentially observable. 
We will also take $\chi_0 = 2\times 10^{-3}$ to generate a significant non-Gaussianity. To see a significantly scale-dependent non-Gaussianity we take the self-interaction with $n=6$ and $\lambda=0.2$, which gives $s_k \approx 0.07$.

\begin{figure}[!h]
\centering
	\begin{subfigure}[b]{0.45\textwidth}
    \includegraphics[width=\textwidth]{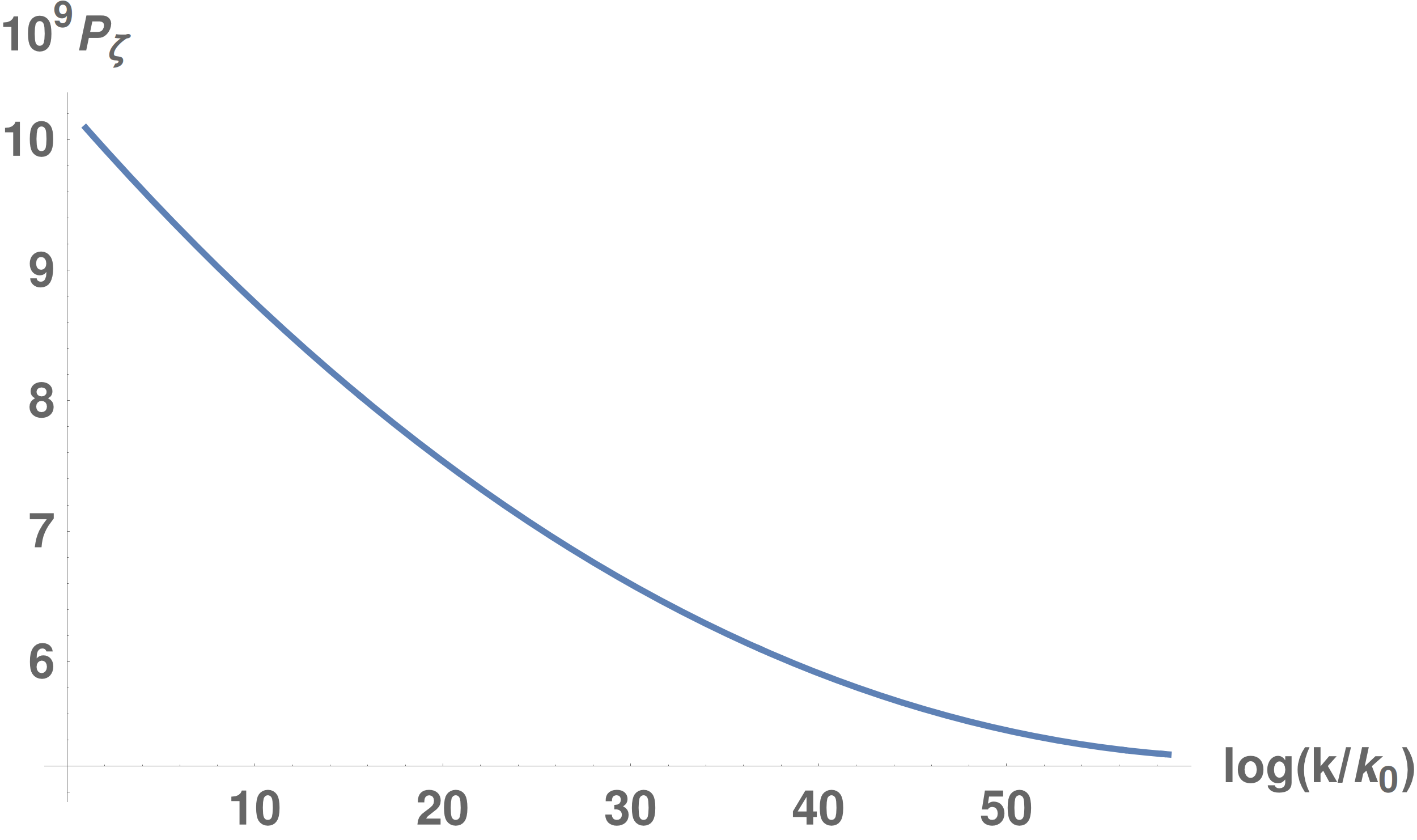}
    \caption{Power spectrum} 
    \label{fig:02_07power2}
    \end{subfigure}
    \begin{subfigure}[b]{0.45\textwidth}
	\includegraphics[width=\textwidth]{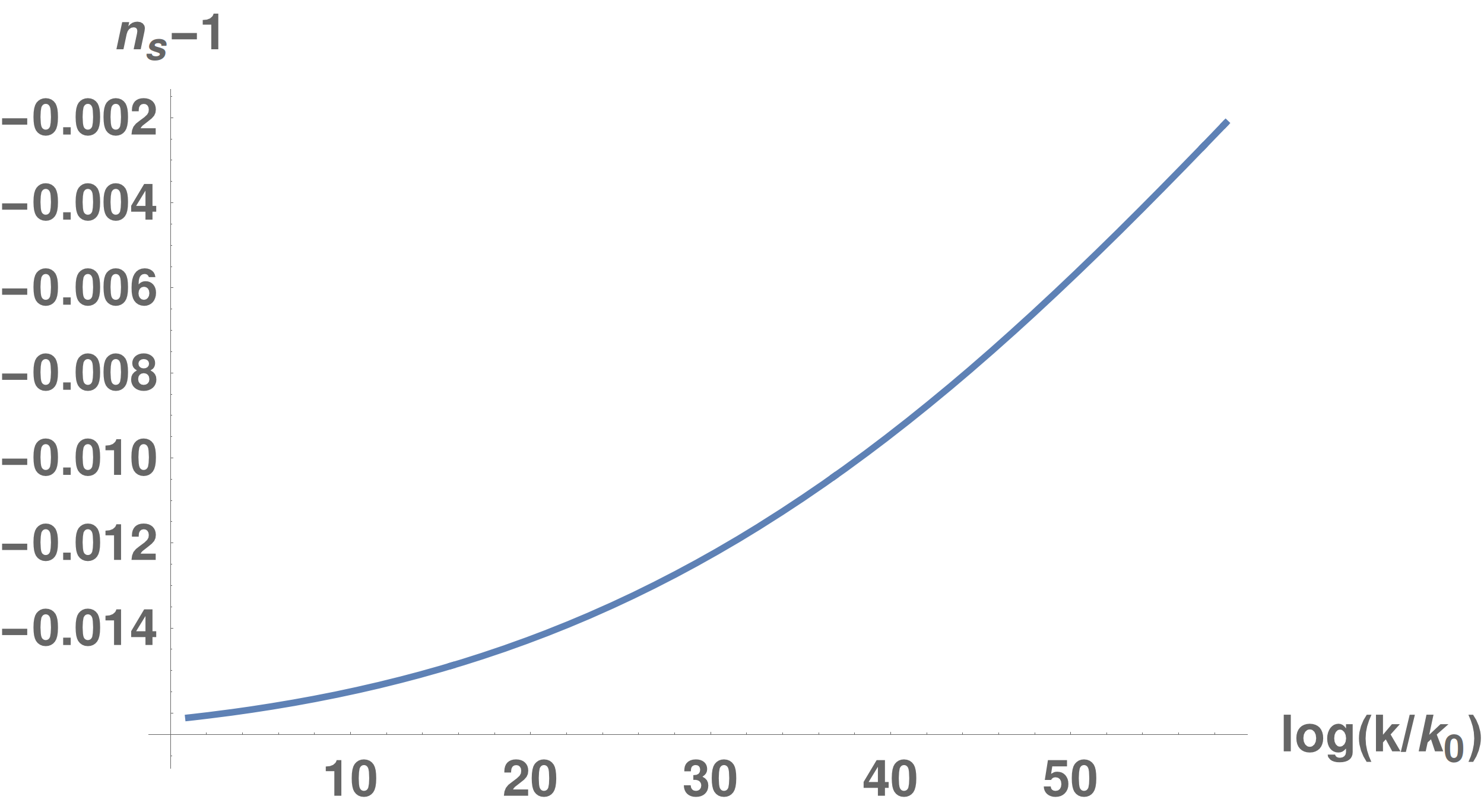}
	\caption{Tilt of the power spectrum} 
	\label{fig:02_07ns}
    \end{subfigure} 
	\begin{subfigure}[b]{0.45\textwidth}
	\includegraphics[width=\textwidth]{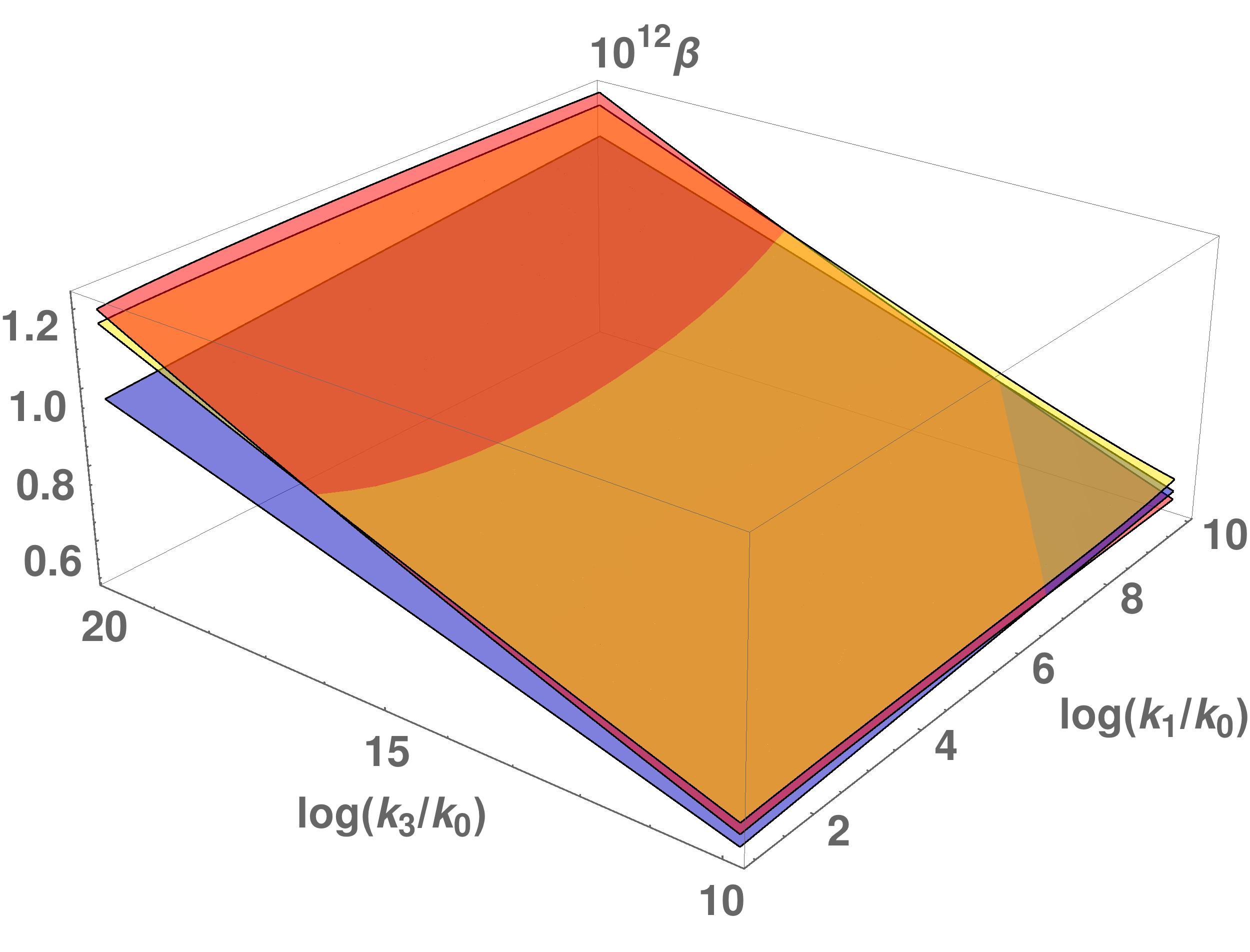}
	\caption{Bispectrum $\beta = 4k_1^3k_3^3B_{\zeta}$ in \protect\eqref{bispectrumsqueezed2}} 
	\label{fig:02_07beta2}
    \end{subfigure}
    	\begin{subfigure}[b]{0.45\textwidth}
	\includegraphics[width=\textwidth]{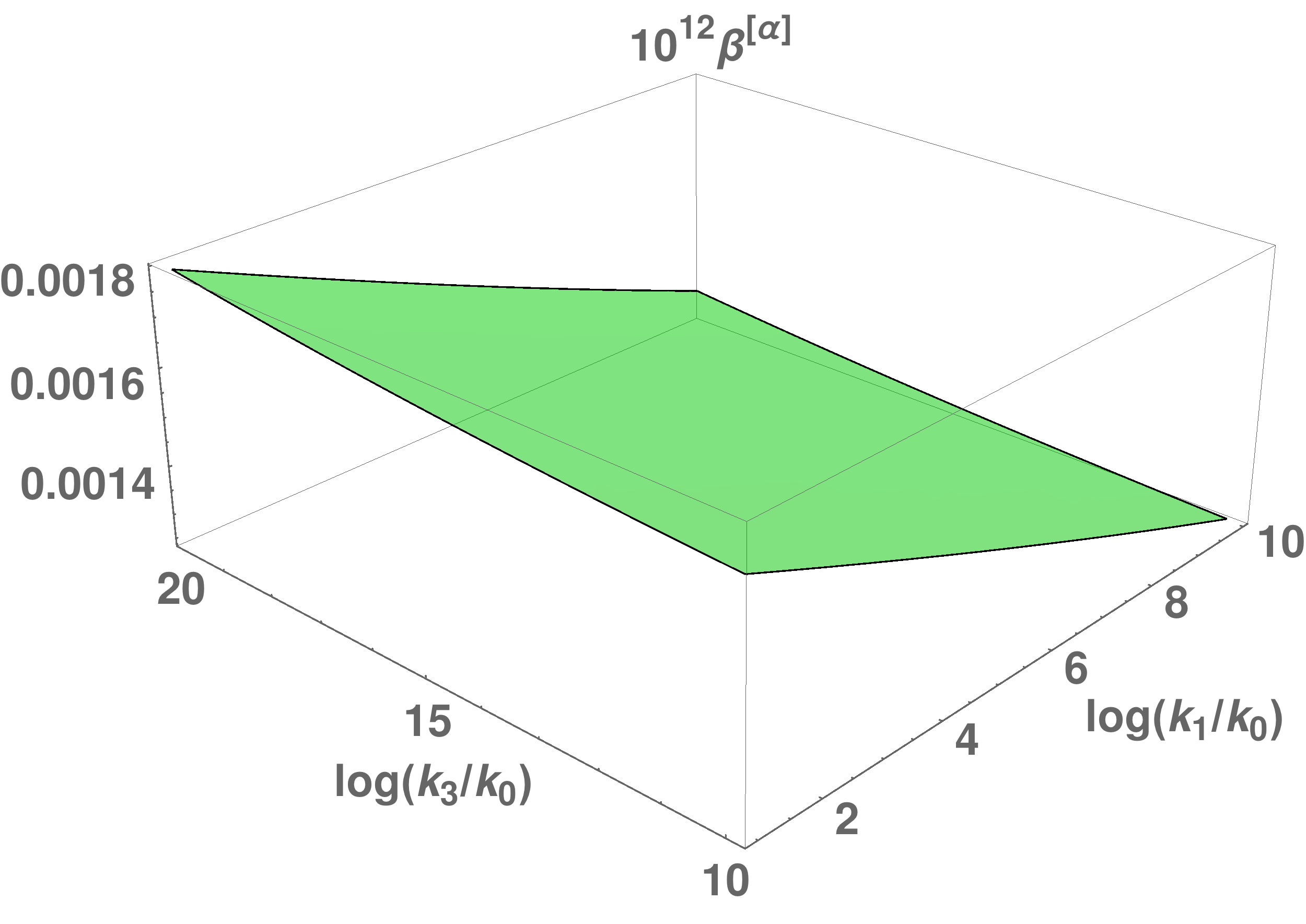}
	\caption{$\beta^{[\alpha]} = 4k_1^3k_3^3B_{\zeta}^{[\alpha]}$ from first line of \protect\eqref{bispectrumsqueezed2}} 
	\label{fig:02_07alpha}
    \end{subfigure}
    	\begin{subfigure}[b]{0.45\textwidth}
	\includegraphics[width=\textwidth]{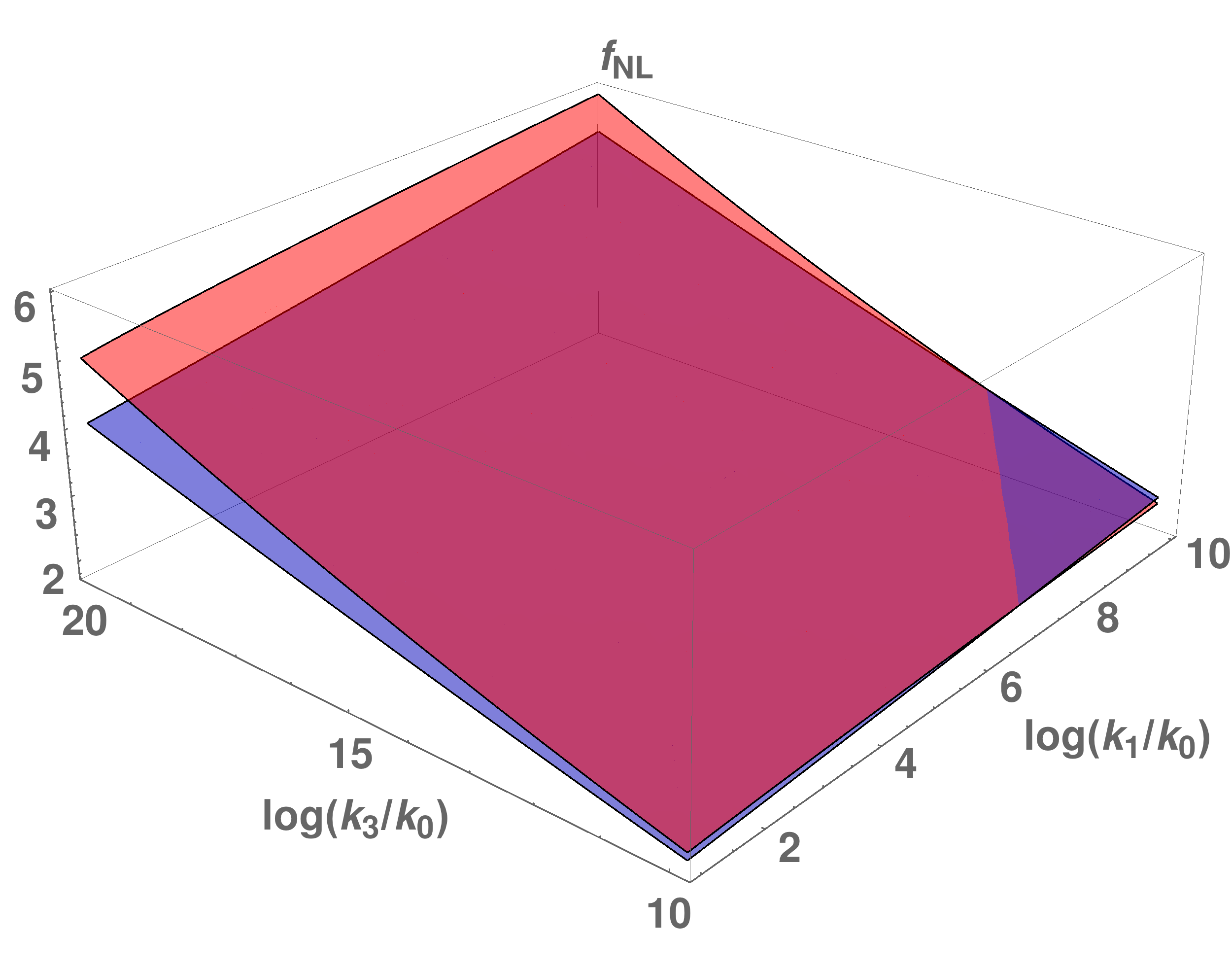}
   \caption{Reduced bispectrum $\fnl$, \protect\eqref{fnl}} 
   \label{fig:02_07fnl2}
    \end{subfigure}
    \begin{subfigure}[b]{0.45\textwidth}
	\includegraphics[width=\textwidth]{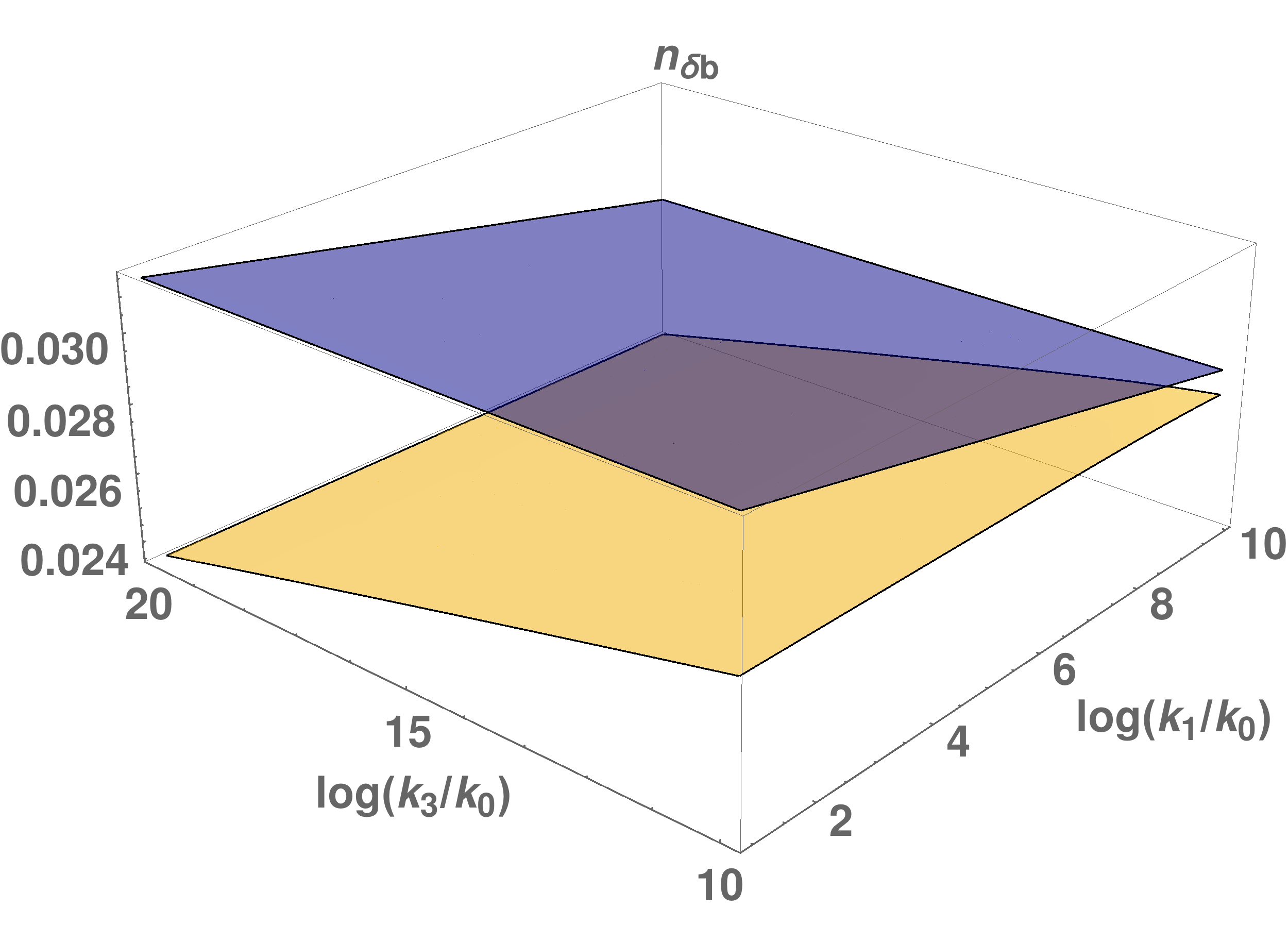}
	\caption{Spectral index of the halo bias, \protect\eqref{ndbsimple}} 
	\label{fig:02_07ndb2}
    \end{subfigure}     
\caption{
In the 3d plots, the blue surfaces are our expressions with $\Gamma$'s. Red surfaces are Byrnes {\it et al.} \protect\cite{2010JCAP...02..034B} expressions. Yellow surfaces are the Dias {\it et al.} \protect\cite{Dias:2013rla} expressions. The near-equilateral regime is on the far right of each plot. The highly squeezed limit is on the far left of each plot. The green surface shows the $[\alpha]$ contribution is negligible for this model.
} 
\label{fig:3dplots}
\end{figure}


The power spectrum for this model is shown in Figure~\ref{fig:02_07power2}, and the tilt of the power spectrum is shown in Figure~\ref{fig:02_07ns}. In order to see a large scale dependence of $\fnl$ for this simple model, we took the curvaton mass to be as large as the inflaton mass\footnote{
A potential issue for the mixed curvaton-inflaton model with equal masses is that since both fields begin oscillating at exactly the same time, the curvaton energy density remains subdominant for a long time -- see \cite{Hardwick:2015tma} for a discussion on this. }. However this has the adverse effect of making the tilt of the power spectrum to not be as red (negative) as observations suggest. This could be fixed by beginning with a more complicated model, and not requiring the curvaton mass to be as large, but this is not our focus here, and instead we focus on the results for the bispectrum. 

We can now compare our new expressions with $\Gamma$'s against existing expressions in the literature. In the 3d plots, the near-equilateral limit $\log(k_1/k_3) \approx 0$ is on the far right of each plot, and the squeezing increases as you move away from this corner. The highly squeezed limit, where $\log(k_1/k_3) \sim -20$ is on the far left of each plot.

In Figure~\ref{fig:02_07beta2} we plot the bispectrum $\beta \equiv 4k_1^3k_3^3B_{\zeta}$ to show how the bispectrum scales with $\log( k_1/k_0)$ and $\log( k_3/k_0) $ where $k_0=H^{(0)}$ at the initial time $t_0$, where we have set $a_0=1$. The blue surface uses our new expression \eqref{bispectrumsqueezed2}, the red surface uses (96) of Byrnes {\it et al}. \cite{2010JCAP...02..034B} given in 
\eqref{BNTW96}, and the yellow surface uses (14) of Dias {\it et al.} 
\cite{Dias:2013rla}, given in \eqref{DRSbis}. In the highly squeezed limit we see a percentage difference of about $20\%$. The near-equilateral results are within a few percent of each other for each of the three expressions -- the small difference is because previous authors have included an additional constant term which appears at next order in slow-roll in their 
expression for $\Sigma$ at horizon crossing discussed in Footnote~\ref{foot:sigma} and Appendix~\ref{App:BNTW}.

In Figure~\ref{fig:02_07alpha} we plot the contribution to the bispectrum 
from the field-space three-point function $\beta^{[\alpha]} = 4k_1^3k_3^3B_{\zeta}^{[\alpha]}$ where $B_{\zeta}^{[\alpha]}$ is the first line of \eqref{bispectrumsqueezed2}. We see that this is a factor of $10^3$ smaller than the total bispectrum for all scales considered, and so all the $[\alpha]$ terms in observables can be neglected for this model. 

In Figure~\ref{fig:02_07fnl2} we plot the reduced bispectrum, where the blue surface is our new expression \eqref{fnlsqueezedorig} and the red surface is (96) of \cite{2010JCAP...02..034B} divided by the factor $2P_{\zeta}(k_1)P_{\zeta}(k_3)$, i.e. the curly brackets of \eqref{BNTW96}. Dias {\it et al.} don't give an explicit expression for the reduced bispectrum, so we don't plot this here. Again, in the highly squeezed limit we see a percentage difference of about $20\%$, and close agreement in the near-equilateral configuration.

Finally, in Figure~\ref{fig:02_07ndb2} we plot the spectral index of the halo bias. Blue is our new expression \eqref{ndbsimple}, and the yellow surface is (17) of \cite{Dias:2013rla}, given in \eqref{DRSnsq}. Byrnes {\it et al}. \cite{2010JCAP...02..034B} don't give an expression for the spectral index of the halo bias so we don't plot this here. Similarly to the other observables we see a percentage difference of about $20\%$ in the highly squeezed limit. We see agreement to within a few percent in the near-equilateral limit, the small discrepancy being due to the Dias {\it et al.} expression being evaluated at the exit time of $k_t \equiv k_1 + k_2+k_3 $ compared to ours being evaluated at $t_3$.

The 20\% level difference for all these observables for a squeezing of $|\log(k_1/k_3)| \sim \mathcal{O}(20)$ can be estimated heuristically as arising from a scale-dependence of the reduced bispectrum of $\mathcal{O}(0.01)$, multiplied by the squeezing.
We thus expect similar levels of discrepancy for any model where the scale-dependence of the reduced bispectrum is of similar order to the scale-dependence of the power spectrum, when considering a squeezing of $|\log(k_1/k_3)| \sim \mathcal{O}(20)$, and that the percentage difference will scale linearly with the scale-dependence of the 
bispectrum.

The expected levels of squeezing for three future experiments were shown in Table~\ref{tab:exp} of the introduction. The results of this section show that inclusion of the effects of field 
evolution can be important when computing the theoretical predictions of a model 
for comparison against observations for large squeezing, even for this simple model.

\section{Conclusion}
In this work we calculated the squeezed limit of the bispectrum of the curvature perturbation for multiple field inflation. Different scales involved in one triangle of 
the bispectrum will exit the horizon at different times, and previous analytic 
expressions have been limited to a mild squeezing where the exit times are roughly equal. Observations can at present probe only a mildly squeezed limit, but future large-scale structure surveys and observations of CMB $\mu$-distortions will be able to probe a highly squeezed limit. It is important, therefore, to have accurate theoretical predictions for this highly squeezed limit in order to ensure the uncertainty in the prediction is less than the uncertainty in the data. For certain models, our results give a correction at a level of 20\% in the highly squeezed limit compared to extrapolating existing expressions, valid only in the mildly squeezed limit, to the highly squeezed limit.

In order to study this highly squeezed limit, we suggested using the 
elegant $\Gamma$ matrix formalism to account for the  evolution on  superhorizon scales of the 
perturbations between exit times.  We also calculated the intrinsic three-point function of the field perturbations, $\alpha$, in the highly squeezed limit for the first time. We did so by appealing to a soft limit argument, previously used in the single field context for the curvature perturbation. Together these
elements allowed us to extend $\delta N$ expressions for the bispectrum of $\zeta$ to account for multiple crossing times. From this expression, we then obtained the reduced bispectrum and the spectral index of the halo bias. Working with a specific model, the mixed inflaton-curvaton scenario with a self-interacting curvaton, we checked the difference in our theoretical prediction, valid for a large squeezing, against existing predictions, valid for a mild squeezing. As would be expected 
we found significant differences, especially in the highly squeezed limit for the cases in which there is significant scale dependence in the reduced bispectrum.

The overall aim of this paper was to provide clarity in how to confront models of inflation 
against observations sensitive to the squeezed limit of the bispectrum. Our results however, could also be useful to check numerical methods in the squeezed limit. From a theoretical and observational 
point of view, soft-limits -- of which the squeezed limit is the simplest example -- are of considerable interest, and in future work we hope to consider soft limits of higher $n$-point correlation functions for multiple field inflation, using a similar approach. For $n>3$, the story can be more interesting than for the bispectrum, where the only soft limit is when a single external momenta becomes small. Firstly, one can consider multiple-soft limits, where more than one momentum becomes smaller than the others. Moreover, one can also consider the collapsed limit, when an internal momentum becomes soft. Observing these higher-point functions may be even harder than for the bispectrum, but it is still important to have theoretical predictions for multiple field inflation to constrain models using observational limits, in particular to search for deviations from single field inflation. 

Finally, we mention that the intrinsic term in the bispectrum coming from $\alpha$ was negligible compared to the other term in the case study presented. Calculating this expression explicitly 
allowed us to determine this, but more work is required to investigate whether this term can 
ever be as large as, or even dominate over, the other contribution.

\section*{Acknowledgements}
We would like to thank Chris Byrnes, Shailee Imrith, David Seery and Raquel Ribeiro for 
taking the time to comment extensively on a draft of this work. ZK is supported by an STFC studentship. 
DJM is supported by a Royal Society University Research Fellowship.

\appendix

\section{Reduction to single field case}\label{App:SF}
We should check that our expression \eqref{bispectrumsqueezed2} for $B_{\zeta}$ reduces to the Maldacena single field squeezed limit result \cite{Maldacena:2002vr} in the case of one slowly rolling scalar field, given by
\begin{align}
&\lim _{k_1 \ll k_2,k_3} B_{\zeta}(k_1,k_2,k_3) 
\approx -(n^{(3)}_{s}-1)P_\zeta(k_1)P_\zeta(k_3) \label{Maldsqueeze}
\end{align}
with $n_s-1 = 2\eta_{V}-6\epsilon_{V}$, and the superscript $(3)$ denoting evaluation at time $t_3$ when $k_3$ exits. Here $\epsilon_{V}\equiv \frac{1}{2}(V'/V)^2$ and $\eta_{V}\equiv V''/V$ are the potential slow-roll parameters.

Thus we begin with a slowly rolling single field $\phi$. In this case we have
\begin{align}
N_{\phi } = -\frac{H}{\dot{\phi}} = \dfrac{1}{\sqrt{2\epsilon_V}}\,, \qquad \frac{N_{\phi\phi }}{N_{\phi }^2} = 2\epsilon_V - \eta_V\,,
\qquad \frac{d}{d \phi^{(3)}} = \frac{1}{\dot{\phi}^{(3)}}\frac{d}{dt_3}
\end{align}
and the $\Gamma$ matrix is just the number $N_{\phi }^{(1)} /N_{\phi } ^{(3)}$. Then \eqref{bispectrumsqueezed2} simplifies to
\begin{align}
\begin{split}
&\lim _{k_1 \ll k_2,k_3} B_{\zeta}(k_1,k_2,k_3) 
\\
\approx & \frac{(N^{(1)}_{\phi})^2{H^{(1)}}^2}{2k_1^3}\frac{(N^{(3)}_{\phi})^2{{H^{(3)}}}^2}{2k_3^3}\left (-\frac{\dot{\phi^{(3)}}}{{H^{(3)}} }\right )\frac{2}{{H^{(3)}} }\left (\frac{1}{\dot{\phi^{(3)}}}\frac{d {H^{(3)}} }{dt_3}\right )
\\ 
&+ 2 \frac{(N^{(1)}_{\phi})^2{H^{(1)}}^2}{2k_1^3}\frac{(N^{(3)}_{\phi})^2{{H^{(3)}}}^2}{2k_3^3}\frac{N^{(3)}_{\phi\phi}}{(N^{(3)}_{\phi})^2}
\\
\approx & [2\epsilon^{(3)}_{V} + 2(2\epsilon^{(3)}_{V } - \eta^{(3)}_{V })]P_\zeta(k_1)P_\zeta(k_3)
\\
\approx &-(n^{(3)}_{s }-1)P_\zeta(k_1)P_\zeta(k_3) \label{maldrecovered}
\end{split}
\end{align}
as required.

\section{Recovering Seery \& Lidsey result in near-equilateral limit}\label{App:SL}
A further important check is that the near-equilateral limit of our result for $\alpha$, \eqref{squeezealpha} goes over to the result of Seery \& Lidsey \cite{Seery:2005gb}, given earlier in \eqref{SLalpha}, valid when $k_1$ is small, but not so small as to change the exit times appreciably. That is, we want to check that
\begin{align}
\begin{split}
&\lim _{ t_1 \to t_3} N^{(3)}_{i}N^{(3)}_{j}N^{(3)}_{k}\alpha^{(3)}_{ijk}(k_1,k_2,k_3) =
\lim _{k_1 \ll k_2,k_3} \frac{4\pi^4}{k_1^3k_2^3k_3^3}\left (\frac{{H^{(3)}} }{2\pi}\right )^4N^{(3)}_{i}N^{(3)}_{j}N^{(3)}_{k}
\\
&\times \sum_{\text{6 perms}}\frac{\dot{\phi}_i^{(3)}}{4{H^{(3)}}}\delta_{jk}\left (-3\frac{k_2^2k_3^2}{k_t} - \frac{k_2^2k_3^2}{k_t^2}\left (k_1+2k_3\right ) +\frac{1}{2}k_1^2 - k_1k_2^2 \right ).
 \end{split} \label{checkSL}
\end{align}

Beginning with the RHS, one can do the sum over all six permutations, then take the slightly squeezed limit to get the RHS equal to
\begin{align}
\begin{split}
- N^{(3)}_{i}N^{(3)}_{j}N^{(3)}_{j}\frac{{H^{(3)}} ^4}{4k_1^3 k_3^3}\frac{d \phi^{(3)}_i}{d N}
\end{split} \label{SLsq}
\end{align}
which is exactly the $t_1\to t_3$ limit of $\alpha^{(3)}_{ijk} $ in \eqref{squeezealpha}, in which $\Gamma^{(3,1)}_{ij}\to \delta_{ij}$, contracted with $N^{(3)}_{i}N^{(3)}_{j}N^{(3)}_{k}$.

\section{Squeezed limits of graviton correlators} \label{app:grav}
Three-point functions involving gravitons (tensors) are likely significantly harder to detect observationally 
than those just involving scalars. Nonetheless they are interesting to calculate 
with a view to observations in the more distant future, and from a theoretical perspective.

Maldacena found squeezed limits of scalar-graviton and graviton-graviton three-point functions in the case of a single scalar field \cite{Maldacena:2002vr}\footnote{Maldacena and Pimentel \cite{Maldacena:2011nz} found graviton-graviton correlators for gravity theories not restricted to Einstein gravity, using a de Sitter approximation. Squeezed limits of correlation functions involving gravitons in models of quasi single field inflation were considered in \cite{Dimastrogiovanni:2015pla}}. Here we use our soft-limit argument to calculate these in the multiple field case. 

First order graviton perturbations, denoted by $\gamma$, are gauge invariant in contrast to scalar perturbations. They are defined as the transverse traceless perturbations of the spatial metric, $h_{IJ}$, such that
\begin{align}
h_{IJ} = a^2(t)\left [(1+2\zeta)\delta_{IJ} + \gamma_{IJ} \right ]
\end{align}
where $\gamma_{II}=0= \partial_I\gamma_{IJ}$. We can Fourier expand $\gamma$ as
\begin{align}
\gamma_{IJ} = \int \frac{d^3k}{(2\pi)^3} \sum_{s=\pm}\epsilon^s_{IJ}(k)\gamma^s_{\vect{k}}(t) e^{i\vect{k} \cdot \vect{x}}
\end{align}
where $s,r$ indices label the polarization of the graviton, and the polarization tensors $\epsilon^s_{IJ}(k)$ satisfy $\epsilon^s_{II}(k)=0= k_I\epsilon^s_{IJ}(k)$ and $\epsilon^s_{IJ}(k)\epsilon^{r}_{IJ}(k)=2\delta_{sr}$.
The two-point function of the graviton is given by
\begin{align}
\langle \gamma^s_{\vect{k_1}}\gamma^r_{\vect{k_2}} \rangle &= (2 \pi)^3  \delta(\vect{k_1} + \vect{k_2} )P_{\gamma}^{sr(1)}(k_1)
\\
P_{\gamma}^{sr(1)}(k_1) &= \delta_{sr}\frac{2{H^{(1)}}^2}{2k_1^3}.
\end{align}

We now consider the squeezed limit of three-point correlation functions involving gravitons. When the soft mode $k_1$ is that of a $\zeta$, there will be a correlation between $\zeta_{\vect{k_1}}$ and two $\gamma$'s by way of a similar soft limit argument applied now to a $\zeta\gamma\gamma$ correlator, giving the result
\begin{align}
\lim _{k_1 \ll k_2,k_3} \langle \zeta_{\vect{k_1}}\gamma^s_{\vect{k_2}}\gamma^r_{\vect{k_3}} \rangle &=  (2 \pi)^3  \delta(\vect{k_1} + \vect{k_2} + \vect{k_3})
N^{(3)}_{i}\Gamma^{(3,1)}_{in}\Gamma^{(3,1)}_{ml}\Sigma_{nl}^{(1)}(k_1)
 P_{\gamma , m}^{sr(3)}(k_3) \label{zetagammagamma}
\end{align}
which can be contrasted with the corresponding single field result given in \cite{Maldacena:2002vr}. 
In the single field case, the result is proportional to the tilt of the graviton power spectrum, providing another consistency relation between observables. Now in the multiple field case, this consistency relation no longer holds, but instead the squeezed limit three-point function is related to how the two-point $\gamma$ correlator depends on the background scalar fields $\phi_i$.

When the soft mode is instead a graviton, we can refer to Maldacena's argument that when the $\zeta_{\vect{k_2}}, \zeta_{\vect{k_3}}$ modes exit, the graviton with momentum $k_1$ exited much earlier and is already frozen, so that fluctuations of $\zeta$ at time $t_3$ will be those in the deformed geometry of the background $\gamma_{\vect{k_1}}$ mode. The main effect of the deformation of the background geometry is to change the  $\delta_{IJ}k_3^{I}k_3^{J} \to \delta_{IJ}k_3^{I}k_3^{J} - \gamma_{IJ}k_3^{I}k_3^{J}$ inside the correlation function of the two $\zeta$'s (equivalently in the second order action for $\zeta_{k_3}$). Putting this into a soft limit type argument gives
\begin{align}
\lim _{k_1 \ll k_2,k_3} \langle \gamma^s_{\vect{k_1}}\zeta_{\vect{k_2}}\zeta_{\vect{k_3}} \rangle &\approx -(2 \pi)^3  \delta(\vect{k_1} + \vect{k_2} + \vect{k_3})P_{\gamma}^{sr(1)}(k_1)\epsilon^r_{IJ}(k_1)k_3^{I}k_3^{J}\frac{d}{d k_3^2}P_{\zeta}(k_3).\label{zetazetagamma}
\end{align}
Similarly, the squeezed limit of the three-point correlator is exactly as given by Maldacena \cite{Maldacena:2002vr}
\begin{align}
\lim _{k_1 \ll k_2,k_3} \langle \gamma^{s}_{\vect{k_1}}\gamma^{r}_{\vect{k_2}}\gamma^{t}_{\vect{k_3}} \rangle &\approx -(2 \pi)^3  \delta(\vect{k_1} + \vect{k_2} + \vect{k_3})P_{\gamma}^{sq(1)}(k_1)\epsilon^q_{IJ}(k_1)k_3^{I}k_3^{J}\frac{d}{d k_3^2}P_{\gamma}^{rt(3)}(k_3).
\end{align}
We highlight that the result \eqref{zetagammagamma} may lead to new shape dependence compared to the single field results.

\section{Reduction to Byrnes {\it et al.}} \label{App:BNTW}
It is also important to 
check our expressions match those which have previously appeared in the literature 
in the near-equilateral, midly-squeezed configuration, which is the overlapping regime of validity. A result for the squeezed limit of the bispectrum was given by 
Byrnes {\it et al.}  in Eq.~(96) of Ref~\cite{2010JCAP...02..034B}
\begin{align}
\begin{split}
B_{\zeta}(k_1,k_2,k_3) = &\left \{\frac{N_{i}^{(1)}N_{j}^{(1)}N_{lm}^{(3)}\left [\delta_{il}+\left  (2c+\log\frac{k_3}{k_1}\right )u_{il}\right ]\left [\delta_{jm}+\left  (2c+\log\frac{k_3}{k_2}\right )u_{jm}\right ]}{N_{r}^{(1)}N_{s}^{(1)}N_{t}^{(3)}N_{z}^{(3)}\left (\delta_{rs}+ 2cu_{rs}\right )\left (\delta_{tz}+ 2cu_{tz}\right )}\right \}
\\
&\times P_{\zeta}(k_1)P_{\zeta}(k_2) + 2\text{ perms} \label{BNTW96}
\end{split}
\end{align}
where $c=2-\log 2 - \gamma$, with $\gamma$ the Euler-Masheroni constant, 
and $u_{ij}$ is given by \eqref{udef}, and since $u_{ij}$ is of order slow-roll, it can be evaluated at any time. Their 
result is valid for small $\log(k_3/k_1)u_{ij}$ with 
the intrinsic contribution from the three-point function of field perturbations 
neglected. It was 
derived by considering the two-point  correlation at unequal times, calculated using the 
expression
\begin{align}
\delta \phi^{(3)}_{i,\vect{k}} &= \left [\delta_{ij}+\log\frac{k_3}{k_1}u_{ij}\right ]\delta \phi^{(1)}_{j,\vect{k }}  \label{BNTW93}
\end{align}
which 
is valid for small  $\log(k_3/k_1)u_{ij}$. However, for a 
large squeezing, $\log(k_3/k_1)u_{ij}$ may not be small, even though $u_{ij}$ is of order slow-roll, 
and one will instead need the full $\Gamma$
expression \eqref{eq:gammaDef} for the evolution of the field perturbations. 
The $\Gamma$ matrix can be written formally as a time-ordered exponential \cite{Yokoyama:2007uu,Yokoyama:2007dw,Yokoyama:2008by,2011JCAP...11..005E,Seery:2012vj,Anderson:2012em}
\begin{align}
\Gamma^{(3,1)}_{ij} = T \exp\left [\int_{N_1}^{N_3}u_{ij}(N)d N \right ]
\end{align}
where $N_1$ is the number of e-folds corresponding to time $t_1$ and similarly for $N_3$. 
Note that in the limit of small $\log(k_3/k_1)u_{ij}$, we have, at leading order in 
$\log(k_3/k_1)u_{ij}$
\begin{align}
\Gamma^{(3,1)}_{ij} \approx \delta_{ij} + \log\left (\frac{k_3}{k_1}\right )u^{(1)}_{ij  } +...
\label{expandG}
\end{align}
As discussed in Footnote~\ref{foot:sigma}, these authors used a next-order in slow-roll expression for $\Sigma$, 
\begin{align}
\Sigma_{ij}^{(*)}(k_1) \approx \frac{{H^{(*)}}^2}{2 k_1^3}(\delta_{ij}+2cu_{ij}). \label{sigmabeyondsr}
\end{align}
Taking our expression for the squeezed limit of the bispectrum \eqref{bispectrumsqueezed}, and substituting the RHS of \eqref{expandG} for $\Gamma$, and replacing ${H^{(*)}}^2\delta_{ij}/2 k_1^3$ with the RHS of \eqref{sigmabeyondsr},  we recover
\eqref{BNTW96}.

The reason we didn't need to use \eqref{sigmabeyondsr} in the main part of this work was because we were throughout working to leading order in slow-roll, rather than next to leading order in slow-roll. The only time we needed to consider slow-roll terms, such as $u_{ij}$, are when they appear multiplied by $\log(k_3/k_1)$, which can be as large as $\mathcal{O}(20)$, in which case $|\log(k_3/k_1)u_{ij}| \sim 1$. Note that the expansion in Eq~\eqref{expandG} cannot be truncated for $|\log(k_3/k_1)u_{ij}| \sim 1$. In the highly squeezed limit this is why the full expression for $\Gamma$ given in Eq~\eqref{gammaresumm} needs to be used instead of Eq~\eqref{expandG}, even though we can safely neglect the slow-roll correction to the power spectrum in Eq~\eqref{sigmabeyondsr}.

\section{Reduction to Dias {\it et al.}} \label{App:DRSreduction}
Dias {\it et al.} \cite{Dias:2013rla} used a next-to-leading order expression for the 
bispectrum \cite{2013JCAP...10..062D}
\begin{align}
&\lim_{k_1 \ll k_2,k_3}  B_{\zeta}(k_1,k_2,k_3)\approx 2N_{ij }N_{l}N_{m }\left [\Sigma_{il}(k_1)\Sigma_{jm}(k_2)+\Sigma_{il}(k_2)\Sigma_{jm}(k_3)+\Sigma_{il}(k_3)\Sigma_{jm}(k_1)\right ]
\label{DRSbis}
\end{align}
to calculate the spectral index of the halo bias as 
\begin{align}
\begin{split}
& n_{\delta b} \equiv \frac{d \log B_{\zeta}}{d \log k_1 }   = -2\frac{N_{i }N_{j } N_{k } M_{im}\alpha_{mjk}^{\text{LO}}  + N_{ij }N_{k }N_{l }( M_{im}\Sigma_{mk}+ M_{km}\Sigma_{im})\Sigma_{jl}}
 {N_{n }N_{p } N_{q }\alpha_{npq}^{\text{LO}} + 2N_{n }N_{p }N_{q r } \Sigma_{nq}\Sigma_{pr}
}
+2\frac{N_{i }N_{j}M_{ij}}{N_{l}N_{l}}
\end{split}
\label{DRSnsq}
\end{align}
where the right hand sides of both expressions are evaluated at the time $t_t$ when $k_t \equiv k_1 + k_2 + k_3$ crosses the horizon, and where 
\begin{align}
\alpha_{ijk}^{\text{LO}} &= - H^4\frac{\dot{\phi}_{i}}{H}\delta_{jk} = - H^4\frac{V_{,i}}{V}\delta_{jk} \label{drsblo}
\\
\Sigma_{ij}(k_a) &= {H^{(t_t)}}^2\left [\delta_{ij } + 2r_{ij } - 2M_{ij}\log \left ( \frac{2k_a}{k_t}\right)\right ]
\label{drssig}
\\
\text{with } M_{ij} &\equiv \epsilon\delta_{ij} + u_{ij} 
\\
\text{and } r_{ij} &\equiv \epsilon\delta_{ij}(1-\gamma) + u_{ij}(2-\gamma)
\\
\text{where }  u_{ij} &\equiv \frac{V_{,i}V_{,j}}{V^2} - \frac{V_{,ij}}{V}
\end{align}
in which all quantities on the right hand side are again evaluated at $t_t$, and $\gamma$ is the Euler-Masheroni constant. These results can be trusted for a mild hierarchy of scales, where $|\log(k_1/k_3)|$ is of order a few.  
We now check that our expression for $n_{\delta b}$, \eqref{eq:nsqgamma}, can recover the Dias {\it et al.} result, \eqref{DRSnsq}, in the limit where the exit times are very close. 
To do so we begin with the expressions
\begin{align}
\frac{V^{(1)}_{,l}}{V^{(1)}}\Gamma^{(3,1)}_{ik,l}  = -\frac{\dot{\phi}^{(1)}_l}{H^{(1)}}\Gamma^{(3,1)}_{ik,l}  = -\frac{1}{H^{(1)}}\frac{d}{dt_1}\Gamma^{(3,1)}_{ik}  =- \frac{d}{d\log k_1}\Gamma^{(3,1)}_{ik}.
\end{align}
Now assuming we can swap the limit of differentiation with respect to $k_1$ and the limit of taking $t_1 \to t_3$, and using \eqref{expandG} we get
\begin{align}
\lim _{ t_1 \to t_3} \left (   \frac{V^{(1)}_{,l}}{V^{(1)}}\Gamma^{(3,1)}_{ik,l} \right ) &=  - \frac{d}{d\log k_1}\left ( \delta_{ik} + \log\left (\frac{k_3}{k_1}\right )u^{(1)}_{ik  }+...\right ) = u^{(1)}_{ik  }.
\end{align}
Substituting this into $n_{\delta b}$ of \eqref{eq:nsqgamma} and setting $\Gamma^{(3,1)}_{ij}\to \delta_{ij}$ we get
\begin{align}
\begin{split}
\lim _{ t_1 \to t_3} n_{\delta b} = &-2\frac{N^{(3)}_{i}N^{(3)}_{q}(  N^{(3)}_{q}V^{(3)}_{,j} + 6N^{(3)}_{qj}{{H^{(3)}}}^2    ) M^{(1)}_{ij}}{N^{(3)}_{m}N^{(3)}_{r}(  N^{(3)}_{r}V^{(3)}_{,m} + 6N^{(3)}_{rm}{{H^{(3)}}}^2    )}
+2\frac{N^{(3)}_{i}N^{(3)}_{j}M^{(1)}_{ij} }{N^{(3)}_{m}N^{(3)}_{m}} 
\end{split}
\end{align}
which is the of the same form as \eqref{DRSnsq} when \eqref{drsblo} and \eqref{drssig} are substituted in. Note that in their expression everything on the RHS is instead evaluated at exit time of $k_t=k_1+k_2+k_3$, rather than $t_3$, but in the limit where the exit times are very close, this won't affect the result significantly, and we recover their result. 

\section{Tilt of reduced bispectrum in the squeezed limit}\label{app:tilt}
As discussed in \S\ref{sec:tilt} one can study the tilts of the reduced bispectrum, $\fnl$, in the squeezed configuration with respect to any combination of the $k$-modes which it involves. In particular, one 
can calculate how $\fnl$ of \eqref{fnlsqueezedorig} varies with respect to 
$k_1\approx k_2$ or $k_3$, or some combination of them. The dependence 
can be parametrized by 
\begin{align}
n_{\fnl}^{X} \equiv \frac{d \log |\fnl|}{ d \log X}
\end{align}
for $X=k_1,k_3$. In \S\ref{sec:tilt} we found $n_{\fnl}^{k_1}=n_{\delta b}$, 
where $n_{\delta b}$ was calculated in \S\ref{sec:halo} in \eqref{eq:nsqgamma}.
To find $n_{\fnl}^{k_3}$ we write \eqref{fnlsqueezedorig} in a form where the second 
square bracket contains all the $k_3$ dependence
\begin{align}
&\lim _{k_1 \ll k_2,k_3} \frac{6}{5}\fnl(k_1,k_2,k_3) 
\approx \left [\dfrac{1}{N^{(1)}_{q}N^{(1)}_{q}} \right ]\left [ L^{(3,1)}_{ij} \left (N^{(3)}_{i}[\log {H^{(3)}} ]_{,j} + \frac{N^{(3)}_{i}N^{(3)}_{jk}N^{(3)}_{k}}{N^{(3)}_{p}N^{(3)}_{p}}\right )\right ]
\end{align}
so that
\begin{align}
n_{\fnl}^{k_3} \approx & \frac{1}{\fnl}\left [ P^{(3,1)}_{ij,3}\left (N^{(3)}_{i}[\log {H^{(3)}} ]_{,j} + \frac{N^{(3)}_{i}N^{(3)}_{jk}N^{(3)}_{k}}{N^{(3)}_{p}N^{(3)}_{p}}\right ) + L^{(3,1)}_{ij}Q^{(3)}_{ij}                \right ]\label{nfnlk3}
\\
\begin{split}
\text{where } 
Q^{(3)}_{ij} \equiv \  &\frac{1}{2}N^{(3)}_{i}u^{(3)}_{jk}\frac{V_{,k3}}{{V^{(3)}}} - \frac{1}{2}N^{(3)}_{k}u^{(3)}_{ik}\frac{V_{,j}}{{V^{(3)}}} - \frac{N^{(3)}_{i}N^{(3)}_{jk}N^{(3)}_{k}N^{(3)}_{m}N^{(3)}_{n}u^{(3)}_{mn}}{(N^{(3)}_{p}N^{(3)}_{p})^2}
\\
& - \frac{1}{N^{(3)}_{p}N^{(3)}_{p}}\left ( N^{(3)}_{i} N^{(3)}_{k} N^{(3)}_{jkl} \frac{V^{(3)}_{,l}}{{V^{(3)}}} + N^{(3)}_{k}N^{(3)}_{l}N^{(3)}_{jk}u^{(3)}_{il} + N^{(3)}_{i}N^{(3)}_{l}N^{(3)}_{jk}u^{(3)}_{kl}\right )
\end{split}
\\
\text{and } 
P^{(3,1)}_{ij,3} \equiv &\frac{d   L^{(3,1)}_{ij}}{d \log k_3} = - \frac{V^{(3)}_{,l}}{{V^{(3)}}} \Gamma^{(1,3)}_{ml} \left (\Gamma^{(3,1)}_{ik,m} \Gamma^{(3,1)}_{jk} +\Gamma^{(3,1)}_{ik}\Gamma^{(3,1)}_{jk,m} \right ).
\end{align}
Note that we have neglected the intrinsic contribution in Eq.~\eref{fnlsqueezedorig} for simplicity.

To compare with observations, one might wish to use the variables in \cite{Fergusson:2006pr}, given by $\tilde{k},\tilde{\alpha},\tilde{\beta}$, defined as
\begin{align}
\tilde{k}  = \frac{1}{2}k_1 +\frac{1}{2}k_2 + \frac{1}{2}k_3,
\qquad
\tilde{\alpha}  = \frac{k_2-k_3}{\tilde{k} },
\qquad
\tilde{\beta}  = \frac{\tilde{k}-k_1}{\tilde{k} }
\end{align}
which in the squeezed limit are related to $k_1, k_3$ by
\begin{align}
k_1 = \tilde{k}\frac{1-\tilde{\beta}}{3-\tilde{\beta}}, \qquad k_3 = \frac{\tilde{k}}{3-\tilde{\beta}} \
\end{align}
with $\tilde{\alpha} \approx 0$.
We can use the chain rule to calculate
\begin{align}
n_{\fnl}^{\tilde{k}} &= \frac{\partial \log \fnl}{\partial \log \tilde{k}} = \frac{\partial  \log \fnl}{\partial  \log k_1}\frac{\partial  \log k_1}{\partial  \log \tilde{k}} + \frac{\partial \log \fnl}{\partial \log k_3}\frac{\partial \log k_3}{\partial \log \tilde{k}} 
= n_{\fnl}^{k_1} + n_{\fnl}^{k_3} \label{nfnltildek}
\\
n_{\fnl}^{\tilde{\beta}}&= \frac{\partial \log \fnl}{\partial \log \tilde{\beta}} = \frac{\partial  \log \fnl}{\partial  \log k_1}\frac{\partial  \log k_1}{\partial  \log \tilde{\beta}} + \frac{\partial \log \fnl}{\partial \log k_3}\frac{\partial \log k_3}{\partial \log \tilde{\beta}} 
= \frac{ -2\tilde{\beta}n_{\fnl}^{k_1}}{(1- \tilde{\beta})(3- \tilde{\beta})} +  \frac{ \tilde{\beta}n_{\fnl}^{k_3}}{(3- \tilde{\beta})}.\label{nfnltildebeta}
\end{align}
Note that in the squeezed limit $\tilde{\beta} \approx 1$, and so if we use our expression for the squeezed limit of $\fnl$ in \eqref{fnlsqueezedorig}, we shouldn't vary $\tilde{\beta}$ significantly away from $1$.

\bibliographystyle{JHEP}

\bibliography{squeeze.bib}

\end{document}